\documentclass{article}

\usepackage{PRIMEarxiv}

\usepackage[utf8]{inputenc} 
\usepackage[T1]{fontenc}    
\usepackage{hyperref}       
\usepackage{url}            
\usepackage{booktabs}       
\usepackage{amsfonts}       
\usepackage{nicefrac}       
\usepackage{microtype}      
\usepackage{lipsum}
\usepackage{physics}
\usepackage{fancyhdr}       
\usepackage{graphicx}       
\usepackage{amsmath}        
\graphicspath{{media/}}     

\usepackage{tabularx}
\usepackage{booktabs}
\usepackage{makecell}
\usepackage{cancel}
\usepackage{subcaption}
\usepackage{comment}
\usepackage{placeins}
\usepackage{algorithm}
\usepackage{algorithmic}
\usepackage{xcolor}
\usepackage{listings}
\newcommand{\rev}[1]{#1}
\title{Ascend to Science: Exploration of AI Chips for Scientific Computing}

\author{
  Weicheng Xue\textsuperscript{*}, Kai Yang\textsuperscript{*}, Yongxiang Liu, Baisong Xu, Dengdong Fan, Xianglin Liu\textsuperscript{\textdagger}, \\ Pengxiang Xu\textsuperscript{\textdagger}, Yonghong Tian\textsuperscript{\textdagger} \\
  Pengcheng Laboratory \\
  Shenzhen, China\\
  \texttt{\{xuewch, yangk, liuyx, xubs, fandd, liuxl, xupx, tianyh\}@pcl.ac.cn} \\
  \textsuperscript{*}Equal contribution. \textsuperscript{\textdagger}Corresponding authors. \\
}

\begin{document}
\maketitle

\begin{abstract}
\rev{The rapid rise of AI-oriented accelerators has reshaped compute systems around low-precision tensor engines, raising a practical question for the HPC community: under what conditions can such hardware support scientific workloads that demand numerical robustness, irregular memory access, and scalability? Using the Ascend 910 NPU series as a representative tensor-centric platform, we characterize precision, execution, and memory-hierarchy bottlenecks that hinder the direct deployment of scientific codes. We then develop and evaluate workload-specific mappings across five application studies---HPL-MxP, LRSVD, SGEMM-cube, PQSim, and SMC-X---combining heterogeneous execution, mixed-precision numerical formulations, precision emulation, hierarchical memory orchestration, and communication--computation overlap. These studies show that AI-native NPUs can achieve numerical robustness, competitive performance, and satisfactory scalability when numerical formulation, execution placement, and data movement are addressed in a coordinated manner. Our results provide a state-of-the-practice case study of how scientific workloads can be adapted to tensor-centric architectures, while distinguishing transferable optimization principles from Ascend-specific implementation details.}
\end{abstract}


\section{Introduction}
\label{sec:intro}

Scientific discovery is increasingly driven by large-scale simulation, numerical modeling, and data-intensive analysis, pushing the demand for computational capability beyond what traditional high-performance computing (HPC) platforms can easily sustain. For decades, CPUs and GPUs have formed the backbone of scientific computing by combining massive parallelism with relatively general programmability. Today, however, the rapid rise of modern AI workloads is reshaping the hardware landscape: emerging compute systems are increasingly built around AI-oriented accelerators that prioritize extreme throughput for low-precision tensor operations~\cite{doi:10.1126/science.adu0801}. This trend raises a fundamental question for the HPC community: can architectures originally optimized for AI also serve as effective platforms for general scientific computing?

Hardware vendors are now prioritizing AI acceleration across both domain-specific and general-purpose processors~\cite{JackNaturePhysics}. Dedicated AI chips such as Google TPU~\cite{9351692}, Huawei Ascend NPU~\cite{Liao2021AscendAS, zuo2025servinglargelanguagemodels}, and AWS Trainium~\cite{10.1145/3698038.3698535} have emerged alongside modern CPUs~\cite{IntelSaphaireRapids} and GPUs~\cite{10070122} that increasingly incorporate tensor engines, wider vector units, high-bandwidth memory (HBM), and bandwidth-oriented cache hierarchies. While these trends are highly effective for AI training and inference, they do not directly align with the algorithmic requirements of many scientific applications, which often demand stronger numerical robustness, more irregular memory access, finer-grained synchronization, and different communication patterns~\cite{JackNaturePhysics}. As flagship computing systems become progressively more expensive and increasingly AI-centric, the scientific computing community will need to adapt important workloads to hardware that was not originally designed for them~\cite{doi:10.1126/science.adu0801}.

Bridging this mismatch requires addressing three fundamental gaps. The first is an \emph{execution mismatch}. AI accelerators excel on dense, regular, high-arithmetic-intensity kernels, but many scientific workloads are dominated by memory-bound, reduction-heavy, or irregular operations whose performance does not scale with tensor throughput alone. The second is a \emph{precision gap}. Scientific computing often relies on FP64 or numerically stable FP32 formulations, whereas modern AI processors derive efficiency from low-precision execution formats such as FP16 and BF16. The third is a \emph{data-movement gap}. Scientific applications frequently depend on multi-level data reuse, irregular control flow, and communication-intensive execution, while AI-oriented architectures typically expose performance through explicitly managed memory hierarchies and tile-centric execution models. Together, these gaps make direct deployment of scientific codes on AI accelerators both nontrivial and, in many cases, ineffective.

In this work, we investigate these challenges on the Ascend 910 series NPUs, a representative family of tensor-centric AI processors built around the DaVinci core architecture. Ascend combines Cube Units for dense tensor execution with Vector Units for more general vector processing, making it a useful platform for examining both the opportunities and the limitations of running scientific workloads on AI-oriented hardware. Rather than proposing a new programming model, we use the existing Bisheng C++ programming interface and runtime stack to study how scientific kernels and applications map onto this architecture in practice. \rev{This work is intended as a state-of-the-practice study rather than a claim that individual optimization primitives such as mixed precision, blocking, gate fusion, or communication overlap are themselves new.} Our goal is to port a few workloads, identify required transformations, and extract empirically grounded lessons for scientific computing on tensor-centric processors.

Our study proceeds from kernels to applications. We first characterize representative scientific kernels to expose the asymmetry between Cube-dominated, compute-bound kernels and Vector-/memory-bound kernels, revealing a fundamental kernel--architecture mismatch for scientific workloads. We then use five representative application studies---HPL-MxP, low-rank SVD (LRSVD), SGEMM-cube, quantum circuit simulation (PQSim), and scalable Monte Carlo (SMC-X)---to show how these gaps can be bridged across scientific workload classes. \rev{The five applications cover dominant, non-exclusive NPU mapping regimes: HPL-MxP combines Cube-friendly dense updates with mixed-precision control and communication overlap; LRSVD uses heterogeneous stage placement for precision-sensitive low-rank linear algebra; SGEMM-cube emulates FP32 dense matrix multiplication through low-precision Cube execution and cache-aware pipelining; PQSim represents bandwidth-dominated state-vector simulation; and SMC-X represents irregular Monte Carlo execution. Their implementations are workload-specific, but share precision-aware numerical reformulation, stage-aware CPU/NPU placement, execution restructuring, and explicit memory/communication orchestration.} Across these five applications, we show that scientific performance depends on precision reformulation, execution restructuring, and explicit data-movement orchestration.

The contributions of this work are threefold:
\begin{itemize}
\item We present a \emph{kernel-centric characterization} of Ascend 910A/910B/910C for scientific computing, showing how representative kernels interact with Cube Units, Vector Units, and the multi-tier memory hierarchy, and identifying the conditions under which scientific workloads can or cannot benefit from tensor-centric execution.

\item We develop \emph{heterogeneous execution strategies} for precision-sensitive scientific applications on AI-oriented architectures, combining stage-specialized CPU--NPU partitioning, numerical stabilization under precision constraints, and \rev{system-level mixed-precision mapping for HPL-MxP}.

\item We derive a set of \emph{systematic optimization principles} for scientific computing on NPUs, including hierarchical memory orchestration, workload restructuring, heterogeneous CPU--NPU mapping, precision emulation, and communication--computation overlap, and validate them across five representative application studies spanning HPL-MxP, LRSVD, SGEMM-cube, PQSim, and SMC-X.
\end{itemize}

The description of artifacts used in this work, including code, benchmarks, and experimental workflows, is provided in Appendix~\ref{artifact_descp}.

\paragraph{Limitations and Scope.}
While our study demonstrates the feasibility of executing representative scientific workloads on AI-oriented Ascend NPUs, several limitations should be acknowledged. First, due to resource and access constraints, most implementations and evaluations in this work are conducted exclusively on the Ascend 910A/910B/910C platforms. 
Although many of the architectural trends we analyze (e.g., tensor-centric compute units, mixed-precision pipelines, and bandwidth-oriented memory hierarchies) are common across modern AI accelerators, our quantitative conclusions may not directly generalize to other NPUs or GPUs without further validation. Second, our methodology relies on carefully designed mixed-precision strategies and kernel-level optimizations to bridge the precision gap between scientific computing and AI hardware. 
The numerical robustness we observe therefore depends on the suitability of these techniques for each workload. 
Applications with stronger sensitivity to round-off error or stricter FP64 requirements may require additional algorithmic reformulation beyond what is explored in this work. Third, our programming and performance results are contingent on the current maturity of the Ascend software stack, including the Bisheng C++ toolchain and runtime. 
Some of the observed performance and programmability trade-offs may evolve as the ecosystem improves, which means that parts of our conclusions are sensitive to the present state of compiler, library, and runtime support. Finally, although we evaluate a diverse benchmark suite spanning micro-kernels, BLAS kernels, and full applications, the selected workloads do not cover the entire spectrum of scientific computing. 
In particular, highly irregular, communication-dominated, or control-flow-heavy applications are underrepresented. 
Extending this study to such workloads and to multi-vendor AI accelerators remains an important direction for future work.


\section{Related Work}

Prior efforts on scientific computing over AI-oriented architectures have made substantial progress, but largely along separate routes. One line of work addresses the \emph{precision gap} between low-precision tensor hardware and the numerical requirements of scientific algorithms. A second line restructures scientific workloads to better match \emph{AI-oriented execution}. A third line focuses on \emph{memory and communication bottlenecks} in large-scale accelerator-based simulations. \rev{We build on these established routes and do not claim that the underlying primitives are new in isolation.} Rather than advancing only one route for one workload class, we connect them into a unified view of scientific computing on AI-oriented NPUs.

\subsection{Precision-Oriented Approaches}

A major challenge in adapting AI accelerators to scientific workloads is preserving numerical robustness on hardware optimized for low-precision execution. Mixed-precision iterative refinement has become a standard solution path in this space. HPL-MxP is a representative example, performing the dominant LU factorization in low precision while recovering FP64-accurate solutions through GMRES-based refinement~\cite{10.1177/10943420251382476, dongarra2024hardware}. This strategy has been successfully deployed on multiple large-scale systems~\cite{kudo2020implementation, lu2022climbing, xue_unlocking}, demonstrating that low-precision accelerators can support numerically reliable scientific solvers when paired with suitable refinement schemes.

Another complementary direction is precision emulation. The Ozaki scheme decomposes high-precision operands into lower-precision segments to emulate more rigorous arithmetic~\cite{ozaki2012error}, while recent tensor-unit implementations of high-accuracy GEMM and related kernels achieve substantial speedups over native high-precision pipelines~\cite{feng2021egemm, ma2022efficiently, xue2025sgemm, 10.1145/3773656.3773670}. These studies provide important numerical techniques, but primarily address precision itself. In contrast, our work treats precision as only one part of a broader architectural mismatch: numerical robustness must be \rev{coordinated} with heterogeneous execution, memory behavior, and workload structure.

\subsection{Workload Restructuring for AI-Oriented Execution}

A second route focuses on reformulating scientific algorithms to better align with the dense, regular, high-throughput operators favored by AI accelerators. Ising model simulations have been reorganized into massively parallel tensor operations~\cite{yang2019high}; density functional theory has been expressed using large-scale tensor contractions~\cite{pederson2023large}; and TPU-based linear algebra frameworks have enabled distributed matrix factorizations and FFTs at scale~\cite{lewis2022large}. Stencil-based models have likewise been cast as matrix multiplications to exploit Tensor Core throughput~\cite{ROMERO2020107473}. JAX-Fluids and JAX-Fluids 2.0 have shown how fluid dynamics can be expressed and differentiated efficiently on AI-oriented systems~\cite{Bezgin2023, Bezgin2025}.

These works show that algorithmic restructuring can unlock AI hardware for specific scientific kernels and applications. However, most focus on one domain or one application family. Our work complements this route by first characterizing representative regular scientific kernels, and then using diverse applications to show how precision-sensitive, bandwidth-dominated, and irregular workloads must be further restructured for efficient execution on AI-oriented NPUs. \rev{The closest delta of our study is therefore not a new blocking or fusion primitive, but the end-to-end mapping of multiple scientific workload classes onto an NPU architecture with explicit Cube/Vector separation, software-managed UB staging, and hierarchical communication.}

\subsection{Memory- and Communication-Centric Approaches}

A third route emphasizes that many scientific workloads on accelerators are constrained less by arithmetic throughput than by memory traffic, data locality, and communication overhead, especially for bandwidth-dominated and irregular workloads. The Cerebras Wafer-Scale Engine, for example, demonstrates extreme-performance molecular dynamics partly by reducing communication bottlenecks through wafer-scale integration~\cite{trifan2022intelligent}. Large-scale mixed-precision solver deployments similarly show that end-to-end efficiency depends on careful coordination among on-chip memory, host-device interaction, and distributed communication~\cite{kudo2020implementation, lu2022climbing, xue_unlocking}.

These studies highlight the importance of data movement as a first-class systems concern, but they typically arise within one benchmark, one architecture, or one application path. More broadly, prior work has largely advanced scientific computing on AI hardware along separate routes: numerical reformulation for low-precision execution, workload restructuring for accelerator-friendly execution, and system-level optimization for memory and communication efficiency. In contrast, our work connects these routes within a unified view of scientific computing on AI-oriented NPUs, combining kernel-level characterization with application-level evidence across HPL-MxP, LRSVD, SGEMM-cube, PQSim, and SMC-X to identify common requirements in numerical reformulation, heterogeneous execution, hierarchical data orchestration, and communication--computation overlap. \rev{We treat this as a practice-oriented synthesis supported by implementation evidence, not as a claim of universal architecture-independent optimality.}

\section{Background and Motivation}

The central question is under what architectural and algorithmic conditions scientific codes can run efficiently and robustly on AI-oriented NPUs. For scientific computing, the answer is governed by three recurring challenges: an \emph{execution mismatch}, because AI processors are optimized for dense, regular tensor operations whereas scientific workloads often contain memory-bound, reduction-heavy, or control-irregular kernels; a \emph{precision gap}, because many scientific applications require numerically stable FP32 or FP64 formulations while NPUs derive efficiency primarily from lower-precision execution; and a \emph{data-movement gap}, because performance is frequently determined not by raw arithmetic throughput alone, but by how effectively computation is orchestrated across on-chip buffers, global memory, and interconnects.

The Ascend 910 series provides a representative platform for studying these issues. Unlike traditional GPGPUs, Ascend was developed as a domain-specific AI processor centered on tensor throughput, software-managed memory movement, and heterogeneous execution engines. These features make it a useful platform for studying the opportunities and limitations of scientific computing on AI-oriented hardware. Our goal is not to argue that NPUs should universally replace general-purpose HPC accelerators, but to determine under what workload and architectural conditions they can serve as effective scientific computing substrates in increasingly AI-centric systems.

\subsection{DaVinci Architecture}

\begin{figure}[h]
  \centering
  \includegraphics[width= 0.8\linewidth, trim=100 15 100 20, clip]{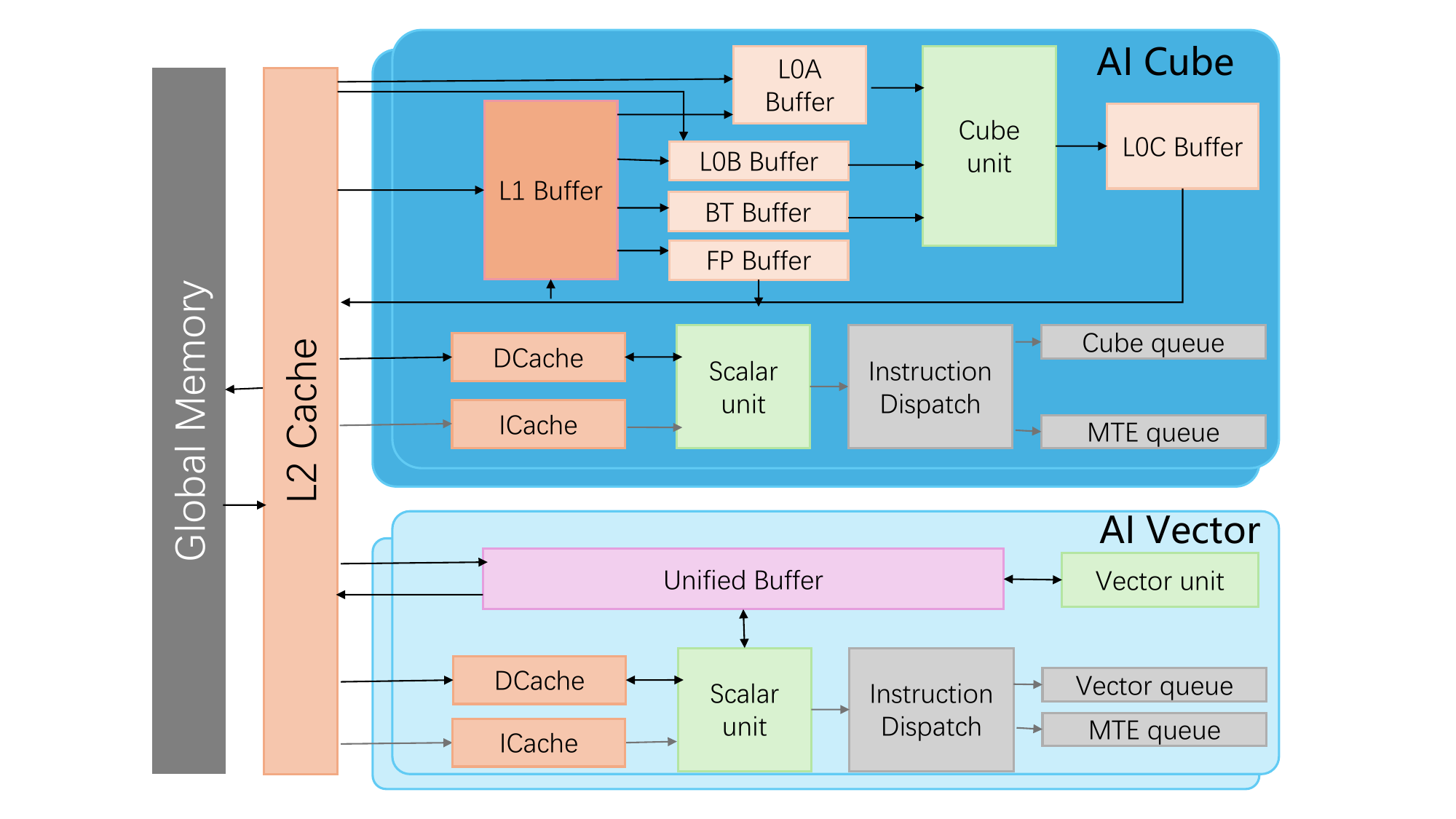}
  \caption{A schematic of the DaVinci architecture. Adapted from~\cite{lin2024fastattention, zuo2025servinglargelanguagemodels}.} \label{fig:NPU_arch}
\end{figure}

The Ascend 910 family is built around the DaVinci architecture, a heterogeneous AI-core design that integrates multiple execution engines with a software-managed memory hierarchy, as illustrated in Fig.~\ref{fig:NPU_arch}. Each AI Core contains both a Cube Core and two Vector Cores, which share access to on-chip storage through explicit data-movement mechanisms. The Cube Core is optimized for dense block-structured computation and operates on fixed-size tensor tiles with dedicated L0/L1 buffers to maximize reuse. The Vector Core, by contrast, provides vector and scalar pipelines with direct access to the Unified Buffer (UB), enabling more flexible element-wise computation, reductions, and control-intensive execution. \rev{The UB is a software-managed on-chip buffer used by Vector/Scalar execution and by MTE-staged data movement. It is distinct from Cube-local L0A/L0B/L0C and L1 buffers, which feed matrix tiles to the Cube pipeline, and from the larger shared L2 buffer that connects AI Cores to off-chip HBM.}

Cube execution offers very high throughput, but only when computation is organized into regular tiles with sufficient arithmetic intensity. Vector execution is more flexible, but provides lower peak throughput and is therefore more exposed to memory-system limitations. At the chip level, AI Cores are connected through a shared L2 buffer and high-bandwidth paths to global memory and HCCS. Data movement among GM, L2, UB, and L0/L1 is explicitly orchestrated by the Memory Transfer Engine (MTE) and associated control logic, making Ascend a compute--memory co-designed architecture whose performance depends on jointly managing execution and data placement.

\subsection{Execution and Memory Implications}

For scientific workloads, the DaVinci core exposes a clear execution asymmetry. Cube-oriented execution is well suited to dense matrix updates, batched contractions, and other kernels with regular access patterns and high arithmetic intensity. Vector-oriented execution, in contrast, is the primary path for kernels dominated by reductions, streaming access, or fine-grained element-wise updates. This distinction implies that scientific kernels on Ascend will not benefit uniformly: some scale with Cube throughput, whereas others remain limited by Vector throughput and memory bandwidth.

This asymmetry is reinforced by Ascend’s multi-tier memory hierarchy (Fig.~\ref{fig:NPU_arch} and Table~\ref{tab:chip_specs}), which spans Cube-local L0/L1 buffers, a shared UB for Vector and Scalar execution, a die-level L2 buffer, and off-chip HBM. Kernels whose working sets can be tiled and staged through on-chip storage can achieve high reuse and approach compute-dominated execution. Dense linear algebra kernels such as GEMM and SYRK fit this model. By contrast, matrix--vector products, reduction-heavy routines, and other weak-locality kernels more readily overflow the on-chip hierarchy and become dominated by traffic among L2, UB, and HBM. For these workloads, explicit orchestration of data placement, buffering, and transfer overlap becomes as important as tensor throughput. These observations motivate our study in this work.

\subsection{Cross-Generation Implications}

As summarized in Table~\ref{tab:chip_specs}, the Ascend 910 family shows a clear cross-generation shift toward a more balanced architecture for numerical workloads. Relative to 910A, later generations improve not only Cube throughput, but also FP32 capability, HBM capacity and bandwidth, and inter-chip communication. These changes matter because they help distinguish gains that come from stronger tensor throughput alone from those enabled by a better balance across compute, memory hierarchy, and communication.

\begin{table}[h]
    \centering
    \caption{Comparison of Ascend 910 Generations. \rev{The 910C peak values are reported per die.}}
    \label{tab:chip_specs}
    \begin{tabular}{l|l|c|c|c}
        \hline
    \makecell[c]{\textbf{Category}} 
  & \makecell[c]{\textbf{Metric}} 
  & \makecell[c]{\textbf{910A}} 
  & \makecell[c]{\textbf{910B}} 
  & \makecell[c]{\textbf{910C}\\\textbf{(per die)}} \\
        \hline
        
Nominal                 & AI Cores (count)           & 32   & 20     & 25 \\
Nominal                 & Vector Cores (count)       & 32   & 40     & 50 \\
FP16 MMA         & Cube FP16 (TFLOPS)         & 256  & 294.9  & 378.9 \\
HF32 MMA         & Cube HF32 (TFLOPS)         & --   & 147.5   & 189.4 \\
FP32 MMA         & Cube FP32 (TFLOPS)         & --   & 73.7   & 94.7 \\
AXPY   & Vector FP32 (TFLOPS)       & 2    & 9.2    & 11.8 \\
Nominal                 & L2 Capacity (MB)     & --   & 192    & 192 \\
L2/UB  & L2 BW (TB/s)        & --   & 7.86   & 7.86 \\
Nominal                 & UB Size (MB)               & 8    & 3.75   & 4.68 \\
Nominal                 & HBM Capacity (GB)          & 32   & 64     & 64 \\
HBM/UB & HBM BW (TB/s)       & 1.2  & 1.6    & 1.6 \\

        \hline
    \end{tabular}
\end{table}

\subsection{Software Stack}

Hardware capability alone does not determine whether AI-oriented processors are useful for scientific computing. Equally important is whether the software stack exposes sufficient control over execution placement, memory movement, and communication to make architectural optimizations accessible in practice. Ascend provides this support through the CANN stack and the Bisheng C++ interface.

CANN (Compute Architecture for Neural Networks) provides the compilation, runtime, and communication support for Ascend NPUs~\cite{liang2020ascend}. Functionally, it plays a role similar to the CUDA toolkit in the GPU ecosystem, integrating device compilation, kernel launch, memory management, and topology-aware collective communication through HCCL. Built on top of CANN, Bisheng C++ provides the low-level heterogeneous programming interface used throughout this work, exposing kernel APIs for AIC/AIV execution, memory and data-movement abstractions, and a queue-based runtime model for asynchronous execution. Together, these layers provide the control needed to express dense Cube-oriented kernels, vector-heavy stages, host--device data movement, and asynchronous execution without introducing a new programming model.

\rev{
Table~\ref{tab:cuda_bisheng_mapping} summarizes the conceptual mapping between
CUDA Tensor Core programming and Bisheng Cube execution. While both
programming models expose accelerator-oriented matrix operations, Bisheng
requires more explicit control over execution placement and data movement
across the software-managed memory hierarchy. This difference explains why
efficient scientific workloads on Ascend often require architecture-aware
kernel restructuring rather than direct source-level translation.
}

\begin{table}[t]
\centering
\caption{\rev{Programming-model mapping between CUDA Tensor Core and Bisheng Cube execution.}}
\label{tab:cuda_bisheng_mapping}
\scriptsize
\begin{tabular}{l|l}
\hline
\textbf{CUDA Tensor Core} & \textbf{Bisheng Cube Execution}\\
\hline
Thread block & AI Core \\
Shared memory & UB/L0 buffer \\
WMMA tile loading & GM$\rightarrow$L0A/L0B data movement \\
Tensor Core MMA & Cube MMA (\texttt{Mmad}) \\
Result write-back & L0C$\rightarrow$GM transfer \\
\hline
\end{tabular}
\end{table}

\subsubsection{Additional Software-Stack Detail}

\begin{figure}[t]
    \centering
    \includegraphics[width=0.8\linewidth, trim=90 428 150 142, clip]{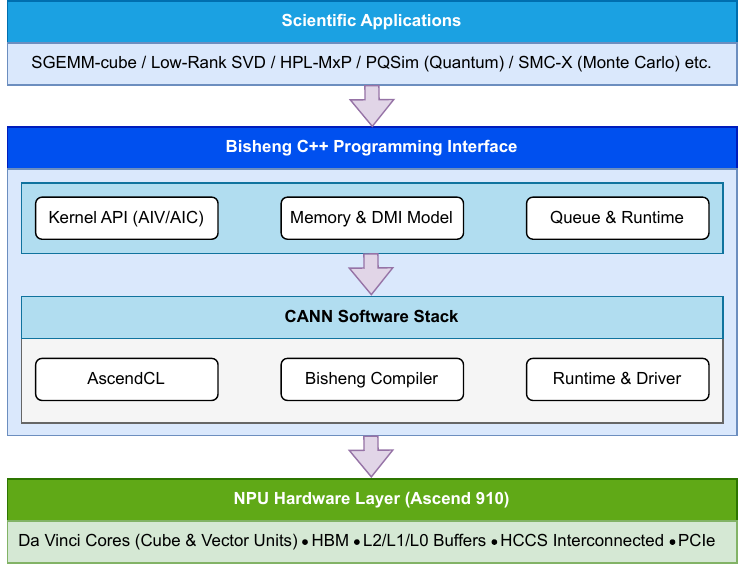}
    \caption{A layered view of the Ascend software stack.}
    \label{fig:ascend_stack}
\end{figure}

\subsubsection{CANN}

CANN (Compute Architecture for Neural Networks) provides the foundational compilation, runtime, and communication support for Ascend NPUs~\cite{liang2020ascend}. It plays a role analogous to the CUDA Toolkit in the GPU ecosystem, integrating device compilation, kernel launch mechanisms, memory management, and performance-optimized libraries for the DaVinci architecture.

Within CANN, the runtime and driver layers are responsible for managing device contexts, allocating and synchronizing memory, and orchestrating kernel execution on the NPU. The Bisheng compiler translates device code into binaries targeting both Cube and Vector execution units of the DaVinci cores. For multi-chip configurations, CANN includes the Huawei Collective Communication Library (HCCL), which provides topology-aware collective primitives. On 910C systems, HCCL leverages the HCCS interconnect to enable low-latency, high-bandwidth communication without host CPU involvement, which is critical for scalable scientific workloads.

\subsubsection{Bisheng C++ Programming Interface}

Built on top of CANN, Bisheng C++ provides a C++-based heterogeneous programming interface that allows scientific developers to express device kernels and offload computation to Ascend NPUs with direct control on low-level hardware. As shown in Fig.~\ref{fig:ascend_stack}, this interface exposes three core components: a kernel API targeting Ascend Vector and Cube units (AIV/AIC), a memory and data-movement model, and a queue-based execution and runtime abstraction.

At the kernel level, Bisheng C++ offers APIs for writing device-side code that maps onto the DaVinci core's Cube and Vector units. This enables performance-critical kernels, such as dense linear algebra or reduction primitives, to be expressed in a structured C++ form while still exploiting the underlying accelerators. At the memory level, the interface provides abstractions for managing host-device data movement and device-side buffers, allowing scientific codes to control data placement and reuse across kernel invocations. At the execution level, a queue and runtime model is used to schedule kernels and coordinate asynchronous execution on the NPU.

\section{Experimental Setup}

\subsection{System Configuration and Interconnect}

For distributed workloads, performance also depends on host--device coordination, communication, and synchronization. The system hierarchy is therefore directly relevant to the scalability results reported later, especially for communication-sensitive workloads such as HPL-MxP.

\begin{figure}[t]
  \centering
  \includegraphics[width=1.0\linewidth, trim=0 35 0 35, clip]{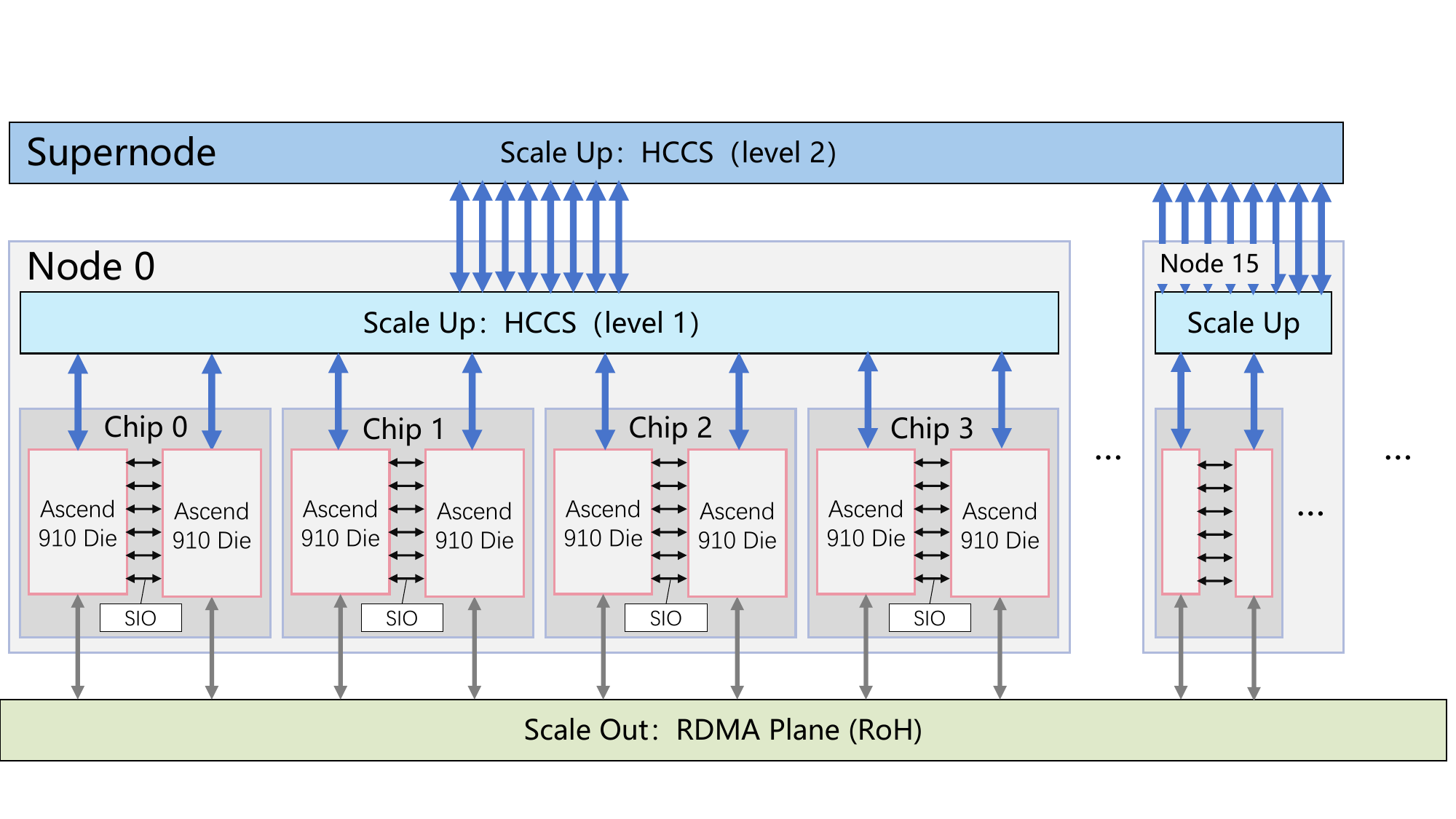} 
  \caption{Hierarchical Interconnect and System Topology of the 910C Cluster. \rev{A standard 910C node contains four physical chips, or eight NPU dies.} Adapted from \cite{zuo2025servinglargelanguagemodels}.} 
  \label{fig:system_hierarchy}
\end{figure}

Each Ascend 910C contains two symmetric dies connected by high-speed SIO links, increasing effective on-package compute density and HBM bandwidth. Within a node, multiple chips are connected through HCCS, which provides a scale-up fabric with substantially higher bandwidth and lower latency than PCIe for collective communication. In our platform, a standard node contains four physical chips, or eight NPU dies, connected through this hierarchy. Beyond a single node, communication extends through HCCS-based scale-up and an RDMA-based scale-out plane (RoH), as illustrated in Fig.~\ref{fig:system_hierarchy}.

In this work, the system hierarchy matters because it determines whether communication-intensive stages can be overlapped effectively with computation and whether multi-chip workloads can scale without excessive synchronization overhead. We therefore treat interconnect behavior as part of the broader data-movement cost of the evaluated system rather than as a secondary implementation detail.

\subsection{Experimental Environment}

To support reproducibility and to clarify platform dependencies, Table~\ref{tab:exp_env} summarizes the hardware and software environments used in this work, including Ascend 910A/910B/910C systems and NVIDIA GPU baselines. Because several results depend on both processor generation and runtime/software-stack support, we report the host CPU and the common software environment for each platform here, and leave application-specific execution settings to the corresponding evaluation subsections.

\begin{table*}[t]
\centering
\caption{Hardware and software environments used in this work. \rev{GPU baselines use optimized vendor libraries where available; application-specific baseline details and timing scopes are described in the corresponding sections.}}
\label{tab:exp_env}
\resizebox{\textwidth}{!}{%
\begin{tabular}{l|l|l}
\hline
\textbf{Platform} & \textbf{Host CPU} & \textbf{Software Environment} \\
\hline
Ascend 910A & 4 $\times$ Kunpeng-920 @ 2.6\,GHz & CANN 6.3.RC2; OpenBLAS 0.3.18; UCX 1.11.2; OpenMPI 5.0.0rc2 \\
Ascend 910B & 4 $\times$ Kunpeng-920 @ 2.6\,GHz & CANN 8.0.RC2.10; OpenBLAS 0.3.18; UCX 1.11.2; OpenMPI 5.0.0rc2 \\
Ascend 910C & 2 $\times$ Kunpeng-920 @ 2.8\,GHz & CANN 8.0.RC2.10; OpenBLAS 0.3.18; UCX 1.11.2; OpenMPI 5.0.0rc2 \\
NVIDIA A800 & 2 $\times$ Intel Sapphire Rapids 8462Y+ @ 2.8\,GHz & CUDA 12.8; cuQuantum 26.01.0 \\
NVIDIA H800 & 2 $\times$ Intel Sapphire Rapids 8462Y+ @ 2.8\,GHz & \rev{CUDA 13.0; cuQuantum 26.06.0} \\
Local Workstation & 2 $\times$ Intel Xeon Platinum 8168 CPU @ 2.7\,GHz & Intel oneAPI HPC Toolkit 2024.2.0 \\ 
\hline
\end{tabular}
}
\vspace{1mm}
\end{table*}

\subsection{Correctness, Variability, and Reproducibility}

All performance results reported in this work were obtained only after validating correctness against application-appropriate reference criteria. For HPL-MxP, correctness follows the standard benchmark requirement that iterative refinement recover an FP64-accurate solution. For LRSVD, correctness was assessed through stable convergence and numerically consistent low-rank decomposition behavior under single-precision constraints. For SGEMM-cube, correctness was assessed through relative error against the FP32 OpenBLAS reference and through stable error behavior across increasing matrix sizes. For PQSim and SMC-X, correctness was verified against the corresponding reference implementations while preserving the intended circuit evolution and Monte Carlo update semantics, respectively.

\rev{For timing, all kernel-level results were measured after warm-up and obtained from at least three repeated runs; we report arithmetic means. Most application-level experiments were repeated three to five times under fixed hardware and software configurations, with additional repetitions when communication jitter was visible. Unless otherwise stated, application timings include the device computation and the communication/data movement required by the evaluated workflow, but exclude one-time compilation and setup costs. The main remaining sources of variability are system-level effects such as communication jitter, runtime scheduling, and host--device synchronization overhead, which are more pronounced in distributed experiments than in single-kernel measurements.}

\rev{To facilitate reproducibility, we report the hardware and software
environments in Table~\ref{tab:exp_env} and describe application-specific mappings and execution strategies in the corresponding sections. Although some optimizations rely on architecture-specific mechanisms such as Cube/Vector execution and software-managed memory, we explicitly describe these mappings to clarify their implementation scope.}

\section{Scientific Kernels}
\label{sec:core_kernels}

The role of this section is not merely to report kernel performance, but to establish a kernel-level basis for the application studies that follow. To understand how scientific workloads interact with Ascend NPUs, we benchmark a representative set of scientific kernels spanning the main regular computational patterns relevant to later applications.

We evaluated a broader set of 23 scientific kernels and selected a representative subset for detailed discussion. Specifically, we choose  dense matrix--matrix multiplication (\texttt{hgemm}, \texttt{sgemm-hf32}) to exercise Cube units under compute-bound conditions, symmetric rank-$k$ update (\texttt{ssyrk}) to represent block linear-algebra updates in factorization algorithms, matrix--vector operations (\texttt{sgemv}, \texttt{ssymv}) to expose Vector execution and memory-bandwidth limits, and complex-valued matrix multiplication (\texttt{cgemm}) to examine a mixed case relevant to quantum simulation. Together, these kernels span different arithmetic intensities, data types, and access patterns. All results are reported as arithmetic means over at least three runs; run-to-run variation is small in these steady-state experiments.

To make the roofline comparison reproducible, Table~\ref{tab:oi_formulas} defines the operational-intensity model used in this section. For each operator, we compute $I=W/Q$, where $W$ is the algorithmic floating-point operation count and $Q$ is the global-memory traffic required to read the inputs and read/write the output; lower-order terms are retained in the formulas. The resulting expressions capture the key separation observed later: the intensity of GEMM-like operators grows with problem dimensions and can become compute-bound, whereas matrix--vector operators approach low, nearly dimension-independent intensity and remain bandwidth-sensitive.

\begin{table*}[t]
\centering
\caption{Operational Intensity (OI) formulations for key BLAS operators on Ascend. Formulas consider theoretical FLOPs and total memory traffic (Read inputs + Read/Write output) including lower-order terms. \textbf{Key:} $M, N, K$ are matrix dimensions. $S$ represents element size in bytes ($S=2$ for FP16, $S=4$ for FP32, $S=8$ for Complex FP32).}
\label{tab:oi_formulas}
\renewcommand{\arraystretch}{1.5}
\resizebox{\textwidth}{!}{%
\begin{tabular}{l|c|c|c|c|c}
\toprule
\textbf{Operator} & \textbf{Precision} & \textbf{Byte Size ($S$)} & \textbf{FLOPs ($W$)} & \textbf{Memory Access ($Q$) [Bytes]} & \textbf{Operational Intensity ($I = W/Q$)} \\
\midrule
\textbf{hgemm} & FP16 & 2 & $2MNK$ & $2(MK + KN + 2MN)$ & $\displaystyle \frac{MNK}{MK + KN + 2MN}$ \\
\hline
\textbf{sgemm} & FP32 & 4 & $2MNK$ & $4(MK + KN + 2MN)$ & $\displaystyle \frac{MNK}{2(MK + KN + 2MN)}$ \\
\hline
\textbf{cgemm} & Complex FP32 & 8 & $8MNK$ & $8(MK + KN + 2MN)$ & $\displaystyle \frac{MNK}{MK + KN + 2MN}$ \\
\hline
\textbf{ssyrk} & FP32 & 4 & $N^2K$ & $4\left(NK + N(N+1)\right)$ & $\displaystyle \frac{N^2K}{4(NK + N^2 + N)}$ \\
\hline
\textbf{sgemv} & FP32 & 4 & $2MN$ & $4(MN + N + 2M)$ & $\displaystyle \frac{MN}{2(MN + N + 2M)}$ \\
\hline
\textbf{ssymv} & FP32 & 4 & $2N^2$ & $4\left(\frac{N(N+1)}{2} + 3N\right)$ & $\displaystyle \frac{N^2}{N(N+1) + 6N}$ \\
\hline
\textbf{ssyr2k} & FP32 & 4 & $2N^2K$ & $4(2NK + N^2 + N)$ & $\displaystyle \frac{N^2K}{2(2NK + N^2 + N)}$ \\ \hline
\textbf{hgemv} & FP16 & 2 & $2MN$ & $2(MN + N + 2M)$ & $\displaystyle \frac{MN}{MN + N + 2M}$ \\ \hline
\textbf{cgemv} & Complex FP32 & 8 & $8MN$ & $8(MN + N + 2M)$ & $\displaystyle \frac{MN}{MN + N + 2M}$ \\ \hline
\textbf{strmv} & FP32 & 4 & $N^2$ & $4\left(\frac{N(N+1)}{2} + 2N\right)$ & $\displaystyle \frac{N^2}{2N^2 + 10N}$ \\ \hline
\textbf{ctrmv} & Complex FP32 & 8 & $4N^2$ & $8\left(\frac{N(N+1)}{2} + 2N\right)$ & $\displaystyle \frac{N^2}{N^2 + 5N}$ \\ \hline
\textbf{batched sgemm} & FP32 & 4 & $2BMNK$ & $4B(MK + KN + 2MN)$ & $\displaystyle \frac{MNK}{2(MK + KN + 2MN)}$ \\
\bottomrule
\end{tabular}%
}
\end{table*}

\begin{figure}[t]
    \centering
    \includegraphics[width=\linewidth]{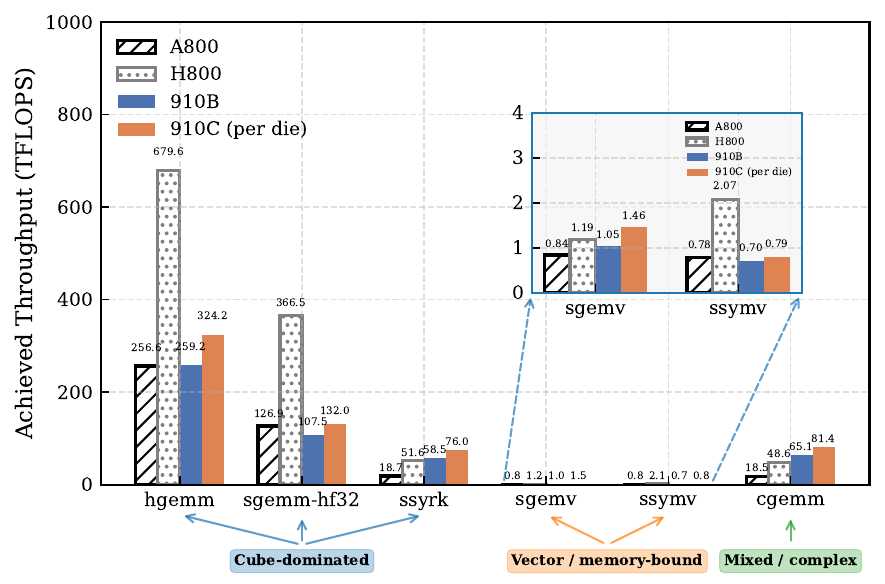}
    \caption{\rev{Achieved throughput of representative scientific kernels on Ascend 910B, 910C, and NVIDIA GPUs.}}
    \label{fig:blas_kernels}
\end{figure}

\begin{figure}[t]
    \centering
    \includegraphics[width=\linewidth]{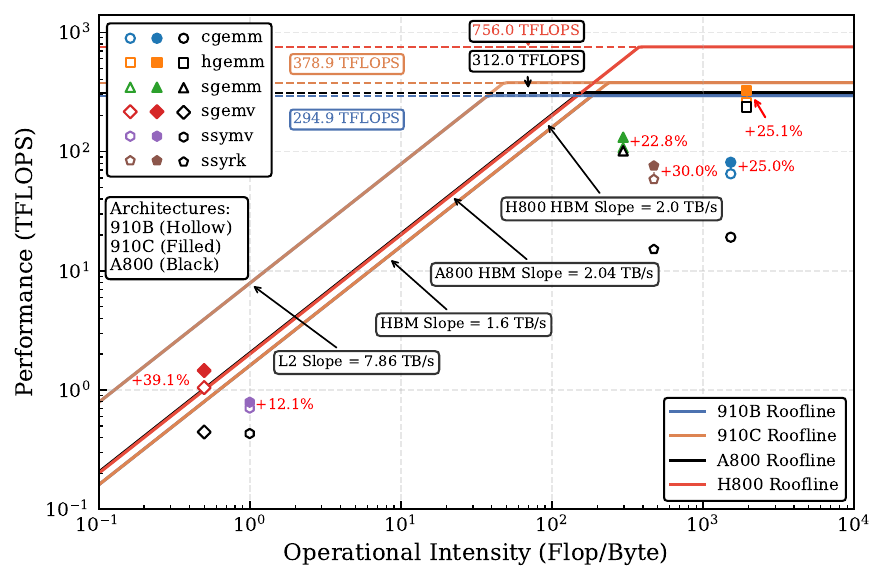}
    \caption{Roofline analysis of representative scientific kernels on Ascend 910B, 910C and GPU. The plot shows achieved performance versus operational intensity (OI) under a single-level global-memory roofline model.}
    \label{fig:kernel_roofline}
\end{figure}

Using the operational intensities defined in Table~\ref{tab:oi_formulas}, Fig.~\ref{fig:blas_kernels} and Fig.~\ref{fig:kernel_roofline} compare the performance of these kernels on Ascend 910B/C and GPU and reveal a clear architectural separation between Cube-dominant and Vector-/memory-dominant execution. For compute-bound kernels such as \texttt{hgemm}, \texttt{sgemm-hf32}, and \texttt{ssyrk}, both processors achieve high fractions of peak tensor throughput, with 910C consistently outperforming 910B. \rev{On 910C, \texttt{hgemm} reaches approximately 324~TFLOPS and \texttt{sgemm-hf32} reaches approximately 132~TFLOPS under the measured device-level configuration. These throughputs correspond to 85.5\% of matching FP16 peak and 69.7\% of matching effective HF32 peak, respectively.} In particular, \texttt{hgemm} reaches approximately 324~TFLOPS on 910C versus 259~TFLOPS on 910B, while \texttt{hgemm} and \texttt{ssyrk} show similar generational gains. These results confirm that dense linear-algebra kernels with regular tiling and high arithmetic intensity map efficiently to the Cube and benefit directly from increased throughput.

In contrast, vector- and memory-bound kernels such as \texttt{sgemv} and \texttt{ssymv} achieve much lower absolute throughput and much smaller generational gains. Their performance remains in the range of 0.7$\sim$1.5~TFLOPS, indicating that these kernels are limited less by available compute peak than by memory bandwidth and Vector-unit efficiency. The modest improvement from 910B to 910C shows that stronger Cube or Vector throughput alone does not substantially accelerate bandwidth-sensitive scientific kernels. Although \texttt{sgemv} appears above the slanted roofline, this does not contradict the roofline model used here, which assumes a single-level global-memory bandwidth. In practice, \texttt{sgemv} benefits from on-chip reuse and Vector--UB pipeline overlap, yielding an effective bandwidth above the simple GM$\rightarrow$L1 assumption.

The complex-valued \texttt{cgemm} kernel occupies an intermediate position. It still benefits from Cube execution, but complex arithmetic and additional data movement reduce utilization relative to real-valued GEMM. \rev{Unlike \texttt{sgemm}'s direct GM$\rightarrow$L1$\rightarrow$L0A/L0B path, interleaved \texttt{cgemm} inputs are first staged in UB for Vector-side real/imaginary unpacking, then moved through L2/L1 to the Cube. This extra unpacking and staging, together with cross-term computation and recombination, lowers utilization. Native complex tile formats or fused real/imaginary unpacking support could reduce this overhead.} As a result, \texttt{cgemm} reaches about 65~TFLOPS on 910B and 81~TFLOPS on 910C: substantially below real-valued GEMM, yet still clearly responsive to improvements in Cube performance across generations.

Taken together, these results establish the conditions under which scientific kernels benefit from Ascend NPUs. High-intensity kernels with regular data access can exploit Cube units effectively and scale with hardware generation, whereas vector-heavy, reduction-dominated, and bandwidth-sensitive kernels remain governed by the Vector engine and memory hierarchy. The application studies that follow can therefore be viewed as different ways of bridging the broader architectural gaps identified earlier: reformulating dense computation into Cube-efficient forms, assigning numerically sensitive or control-intensive stages to more suitable execution paths, and restructuring data movement and communication to reduce the cost of non-Cube-dominant execution.

\section{Application Studies: Bridging Precision, Execution, and Data-Movement Gaps}

The kernel study in Section~\ref{sec:core_kernels} shows that scientific workloads do not benefit uniformly from Ascend NPUs: some map efficiently to the Cube engine, whereas others remain constrained by Vector execution, memory hierarchy, or communication overhead. The five application studies in this section show how the broader precision, execution, and data-movement gaps can be bridged through heterogeneous execution, numerical reformulation, workload restructuring, and hierarchical data orchestration. HPL-MxP spans the precision and communication gaps through mixed-precision refinement and CPU--NPU--communication overlap; LRSVD addresses precision through heterogeneous stage placement and numerical stabilization; and SGEMM-cube addresses precision through FP32 emulation and cache-aware Cube execution. PQSim and SMC-X, by contrast, show how bandwidth-dominated and irregular workloads must be restructured to match the execution and memory model of AI-oriented NPUs.

\subsection{Bridging the Precision Gap}

\subsubsection{Mixed-Precision System-Level Mapping (HPL-MxP)}

\paragraph{Gap and application characteristics}
HPL-MxP has emerged as a representative benchmark for modern supercomputers in the AI era~\cite{10.1177/10943420251382476, haidar2018harnessing}. Unlike traditional FP64-only HPL~\cite{dongarra2003linpack}, HPL-MxP performs the dominant $O(N^3)$ LU factorization in low precision and restores FP64-accurate solutions through $O(N^2)$ iterative refinement. The benchmark thus directly exposes the precision gap between low-precision accelerator throughput and scientific numerical requirements.

On Ascend NPUs, HPL-MxP is challenging not only because of mixed precision, but because it simultaneously stresses dense tensor computation, memory hierarchy, host--device coordination, and distributed communication. The trailing matrix updates are naturally suited to Cube units, but triangular solves and refinement are numerically sensitive and poorly matched to low-precision NPU execution. In distributed runs, panel broadcasts and CPU--NPU coordination introduce additional communication overhead. HPL-MxP thus provides a system-level test of whether mixed-precision scientific computing can be realized efficiently on AI-oriented NPUs.

\begin{figure}[t]
    \centering
    \includegraphics[width=\linewidth,trim=12 12 12 12, clip]{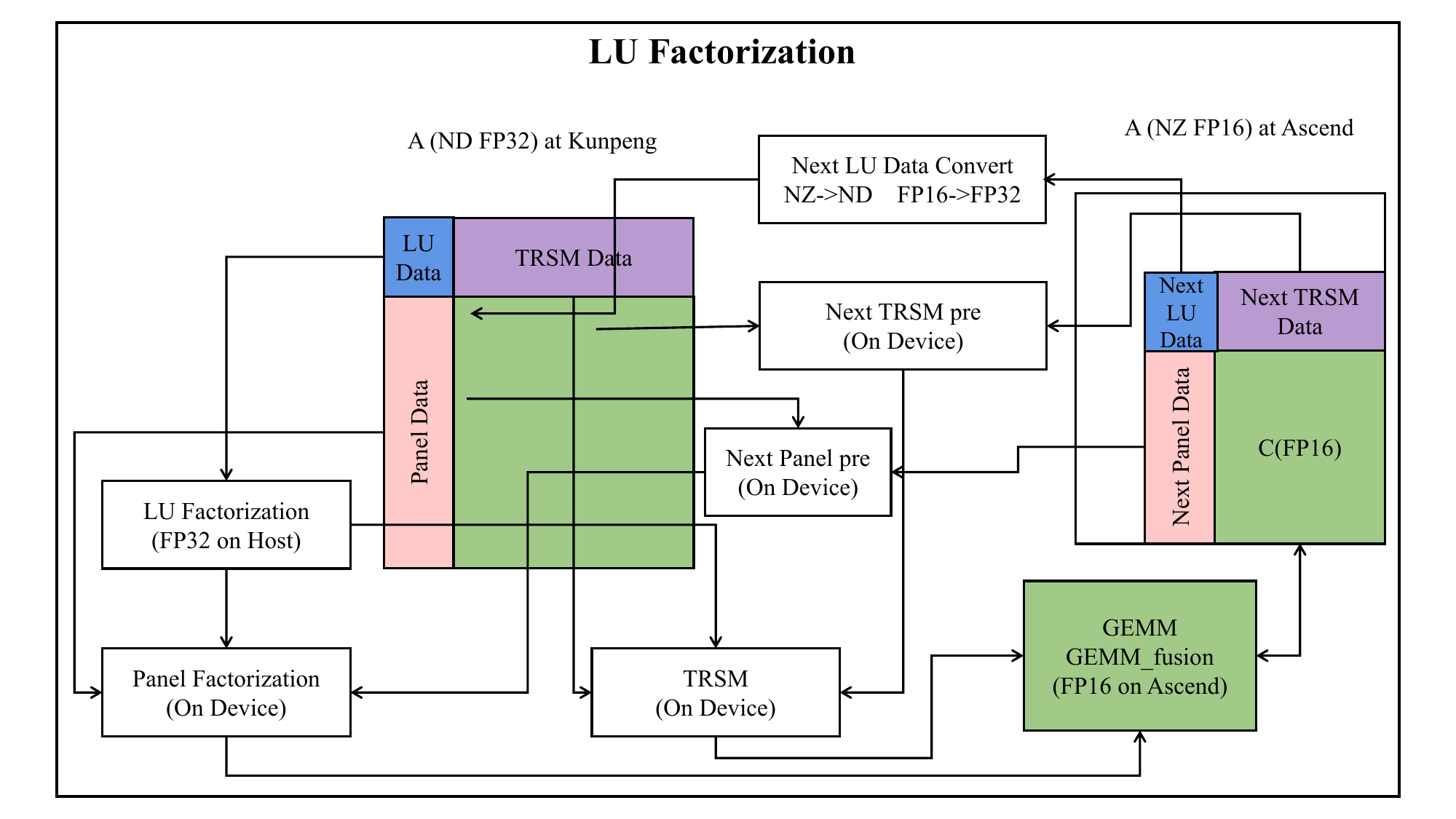}
    \caption{Fused hierarchical pipeline for HPL-MxP LU factorization. The workflow shows orchestration of LU kernels with fused stages and MPI/HCCL communication--computation overlap to maximize NPU utilization. Adapted from~\cite{xue_unlocking}.}
    \label{fig:MxP_workflow}
\end{figure}

\paragraph{Solution: CPU--NPU--communication mapping}
\rev{Our implementation maps HPL-MxP to Ascend through coordinated CPU--NPU--communication mapping (Fig.~\ref{fig:MxP_workflow}).} The computationally dominant trailing matrix update is executed on Ascend Cube units using FP16 inputs with FP32 accumulation, striking a balance between tensor throughput and numerical robustness. To reduce memory pressure and improve reuse, we adopt a multi-iteration fusion strategy that combines multiple GEMM updates into a single execution window, thereby increasing on-chip reuse and overlapping data movement with computation.

Numerically sensitive and control-intensive stages remain on the CPU. In particular, the triangular matrix inversion required by TRSM is carried out in FP64 on the host, and the iterative refinement stage is likewise executed on the CPU because it requires strict FP64 residual computation. This CPU--NPU partition preserves correctness where low-precision NPU execution would be fragile or inefficient.

Communication is handled jointly through HCCL and MPI. Factorized panels are broadcast among NPUs using HCCL, while CPU-side coordination of triangular inverses is handled through MPI. Within TRSM, the CPU computes the inverse of the triangular block, and the resulting matrix is then multiplied with trailing blocks on the NPU Cube units. To integrate these components efficiently, we construct a multi-level pipeline that overlaps computation, memory movement, and communication. Kernel-level pipelining keeps the Cube units saturated, while HCCL and MPI operations are overlapped with NPU execution to reduce idle time at system scale.

\paragraph{\rev{System-level mapping contribution}}
\rev{The contribution here is not the mixed-precision formulation itself, which is prescribed by HPL-MxP, but its system-level implementation on Ascend: dense updates are assigned to Cube units, numerically sensitive stages remain on the CPU, and communication is pipelined with CPU and NPU execution.}

\begin{figure}[t]
    \centering
    \includegraphics[width=\linewidth]{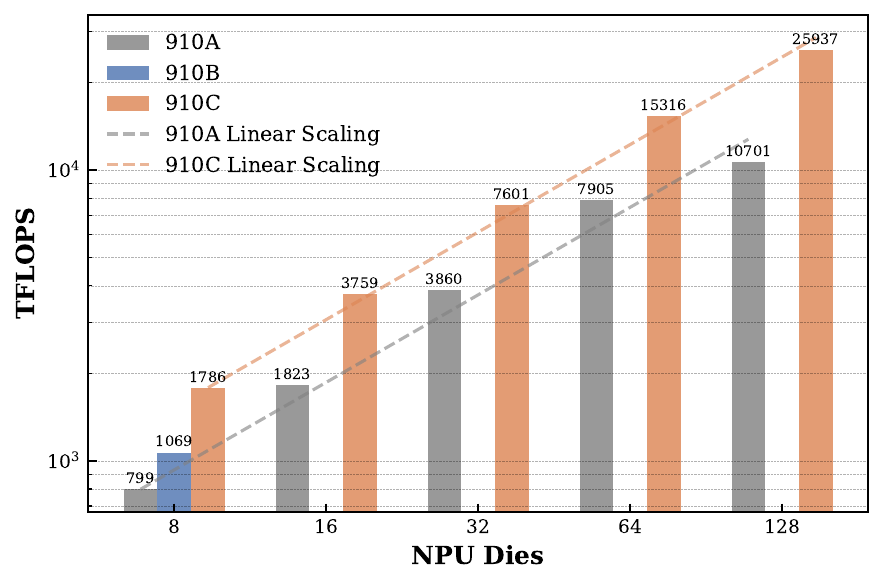}
    \caption{Strong-scaling performance of HPL-MxP on three generations of Ascend NPUs (910A, 910B, and 910C). \rev{The 910B platform is a standalone machine.} The results show sustained TFLOPS across increasing NPU counts, enabled by heterogeneous CPU--NPU mapping,
kernel fusion, and communication--computation overlap.}
    \label{fig:hpl_mxp_perf}
\end{figure}

\paragraph{Validation}
We evaluate the implementation on three generations of Ascend NPUs. Fig.~\ref{fig:hpl_mxp_perf} reports sustained HPL-MxP performance as a function of node count, with each node containing 8 NPU dies. The benchmark exhibits approximately linear scaling over the evaluated configurations, showing that the combined CPU--NPU mapping and multi-level pipeline effectively mitigate both memory and communication bottlenecks. Among the three platforms, Ascend 910C consistently delivers the highest throughput, reflecting improved balance among compute, memory hierarchy, and interconnect.

\rev{Communication is nevertheless a dominant scaling pressure. At 128 dies on the 910A cluster, HCCL collectives account for more than 90\% of the exposed communication time in our breakdown. The multi-level pipeline in Fig.~\ref{fig:MxP_workflow} is therefore essential: it does not eliminate communication, but hides a large fraction of it behind panel preparation, TRSM-related CPU work, and NPU GEMM updates.} These results demonstrate that peak FLOPS alone do not determine scientific performance on AI accelerators. By combining mixed-precision numerical control, fused dense-kernel execution, and aggressive communication--computation overlap, the LU factorization sustains high efficiency without the scalability collapse typical of communication-dominated solvers. This confirms that, for mixed-precision scientific solvers, bridging the precision gap is a system-level problem rather than a kernel-level one: numerical formulation, execution placement, and communication overlap must be \rev{coordinated} for scalable performance on AI-oriented chips.

\subsubsection{Heterogeneous Low-Rank SVD (LRSVD)}

\paragraph{Gap and application characteristics}

Singular Value Decomposition (SVD) is a fundamental primitive in high-performance computing and data analytics, but its cubic complexity makes classical formulations impractical for large-scale problems. Randomized low-rank methods substantially reduce both computational and memory cost while retaining strong accuracy guarantees, making them a natural target for accelerator-based scientific computing~\cite{halko2011finding, musco2015randomized, tropp2019streaming, xu2023fast}.

For Ascend NPUs, the difficulty is not simply that SVD is expensive, but that its phases exhibit sharply different numerical and execution characteristics. Ascend does not provide a native SVD kernel, and its efficient execution path is centered on single-precision tensor and vector operations. As a result, direct deployment of a standard low-rank SVD pipeline leads to both poor hardware utilization and numerical fragility, especially in orthogonalization and subspace construction. LRSVD therefore serves as a representative test case for the precision gap: it contains dense stages that are well matched to Cube/Vector execution, but also numerically sensitive and control-heavy stages that require stabilization beyond straightforward NPU execution.

\begin{figure}[t]
    \centering
    \includegraphics[width=\linewidth]{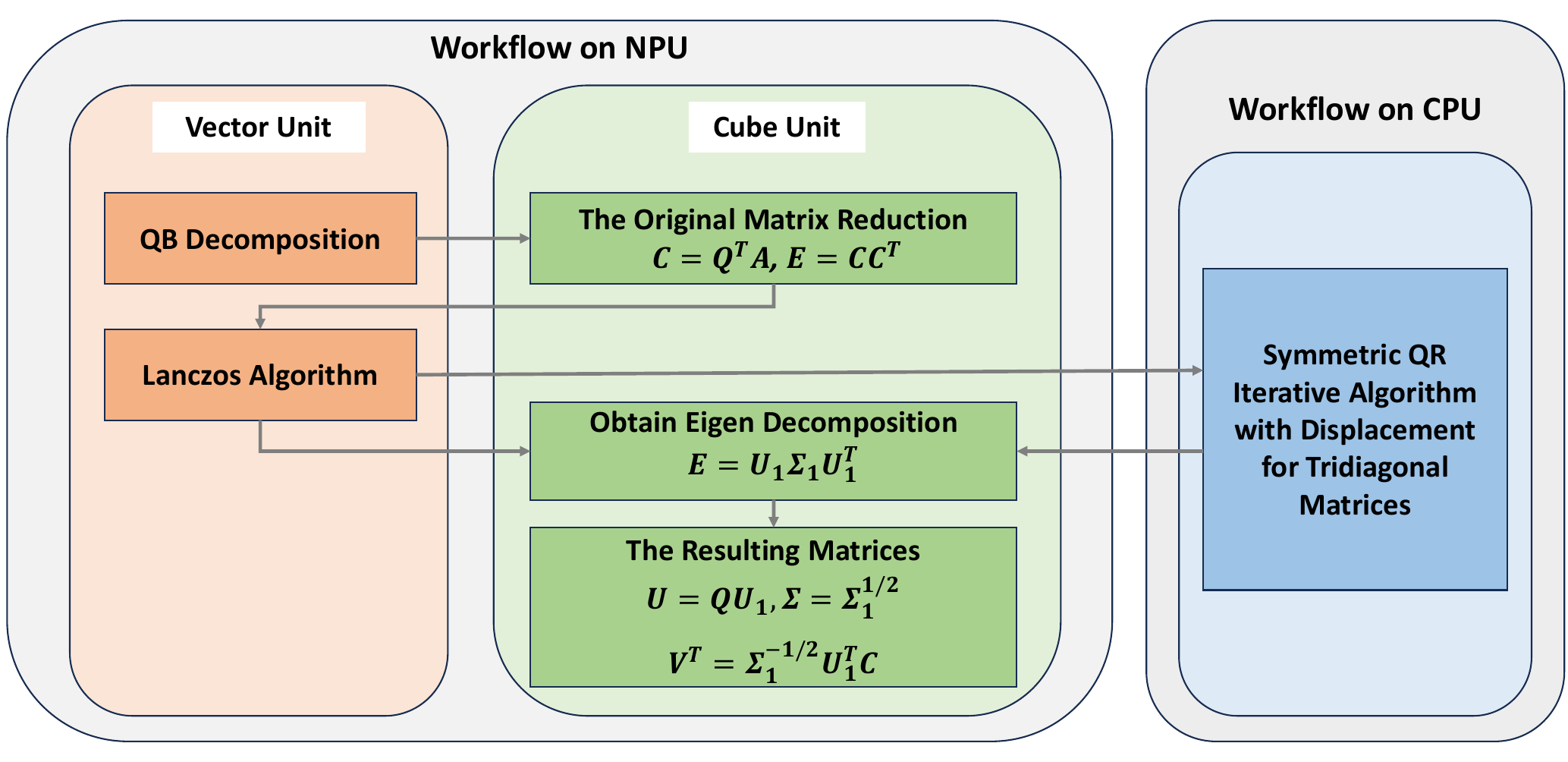}
    \caption{Heterogeneous execution workflow and hardware mapping for the LRSVD solver. The algorithmic pipeline is partitioned by computational structure and numerical sensitivity.}
    \label{fig:lrsvd_workflow}
\end{figure}

\paragraph{Solution: heterogeneous decomposition and refinement}
To bridge this gap, we design a hybrid CPU--NPU low-rank SVD framework that partitions the randomized SVD pipeline by both kernel structure and numerical sensitivity (Fig.~\ref{fig:lrsvd_workflow}). Modern randomized SVD consists of three main phases: (i) QB decomposition, (ii) Lanczos bidiagonalization, and (iii) symmetric QR iteration on a small tridiagonal matrix~\cite{xu2023fast}. These stages map naturally to two execution streams.

The NPU stream targets data-parallel stages dominated by matrix multiplications, projections, and vector norms. QB decomposition and Lanczos iterations are therefore offloaded to the Ascend Cube and Vector units, allowing the computationally intensive sketching and projection phases to exploit high-throughput tensor and vector execution. The CPU stream handles the shifted QR iteration on the small tridiagonal matrix, whose frequent branching and scalar updates make it better suited to the host processor.

\rev{Beyond execution mapping, a second challenge arises from maintaining numerical stability during QB decomposition.} On Ascend, Gram--Schmidt orthogonalization is numerically fragile in single precision: rounding errors accumulate, the orthogonality of basis vectors degrades, and convergence may fail. To restore robustness, we introduce an iterative orthogonalization refinement:
\begin{equation}
    q_{k+1}^{(i)} = \left(I - \sum_{l=1}^{k} q_l q_l^H \right) q_{k+1}^{(i-1)}, \quad i=1,\dots,n_{\text{loop}},
\end{equation}
which reprojects each new basis vector onto the orthogonal complement of the current subspace. In practice, $n_{\text{loop}}=2$ or $3$ provides the best balance between stability and overhead.

\paragraph{\rev{Stage-specialized implementation}}
\rev{The contribution is not randomized low-rank SVD itself, but a stage-specialized heterogeneous workflow for executing LRSVD on AI-oriented NPUs. Data-parallel QB decomposition and Lanczos stages are mapped to Cube/Vector units, while the branch-heavy shifted QR stage remains on the CPU. In addition, single-precision iterative orthogonalization refinement is introduced to recover numerical stability under NPU-oriented execution.}

\paragraph{Algorithmic Optimization for Single Precision}

Directly porting the standard QB decomposition to Ascend's single-precision environment results in significant accuracy loss. We introduce two key modifications to ensure convergence.

\textbf{Iterative Orthogonalization Refinement}\\
The standard update rule for the orthogonal basis vector $q_{k+1}$ is:
\begin{equation}
    q_{k+1} = \left( I - \sum_{i=1}^{k} q_i q_i^H \right) a_{k+j+1}
\end{equation}
In double precision, a single pass of this projection is sufficient. In single precision, residual errors remain in the directions of existing basis vectors $q_1, \dots, q_k$. We replace the single projection with an iterative loop to progressively suppress error accumulation.

\begin{algorithm}
\caption{Refined Basis Calculation with Iterative Orthogonalization}
\begin{algorithmic}[1]
\STATE \textbf{Input:} Current vector $a_{k+j+1}$, existing basis $\{q_1, \dots, q_k\}$, loop count $n_{loop}$
\STATE $q_{k+1}^{(0)} \gets a_{k+j+1}$
\FOR{$i = 1$ to $n_{loop}$}
    \STATE $q_{k+1}^{(i)} \gets \left( I - \sum_{l=1}^{k} q_l q_l^H \right) q_{k+1}^{(i-1)}$
\ENDFOR
\STATE $q_{k+1} \gets q_{k+1}^{(n_{loop})}$
\STATE \textbf{Output:} Refined orthogonal vector $q_{k+1}$
\end{algorithmic}
\end{algorithm}

\begin{figure}[htbp]
     \centering
     \begin{subfigure}[b]{0.48\columnwidth}
         \centering
         \includegraphics[height=3.5cm, width=\textwidth]{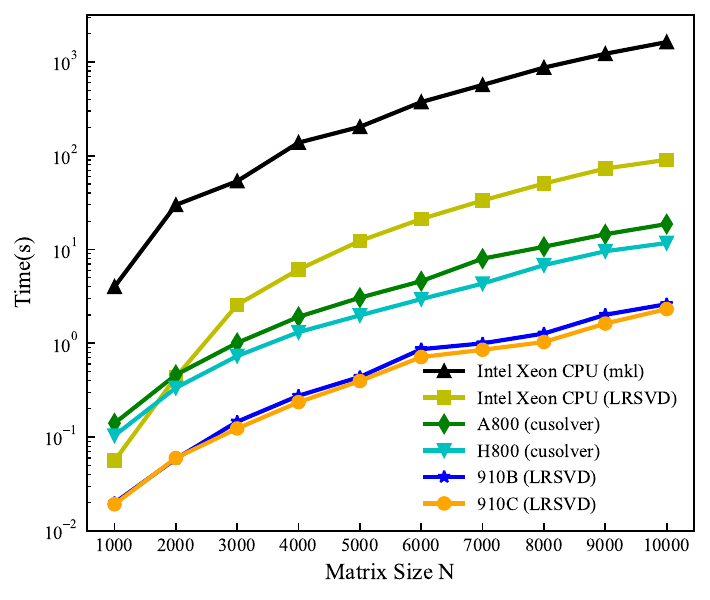}
         \caption{}
         \label{fig:line}
     \end{subfigure}
      \hfill
     \begin{subfigure}[b]{0.48\columnwidth}
         \centering
         \includegraphics[height=3.5cm, width=\textwidth]{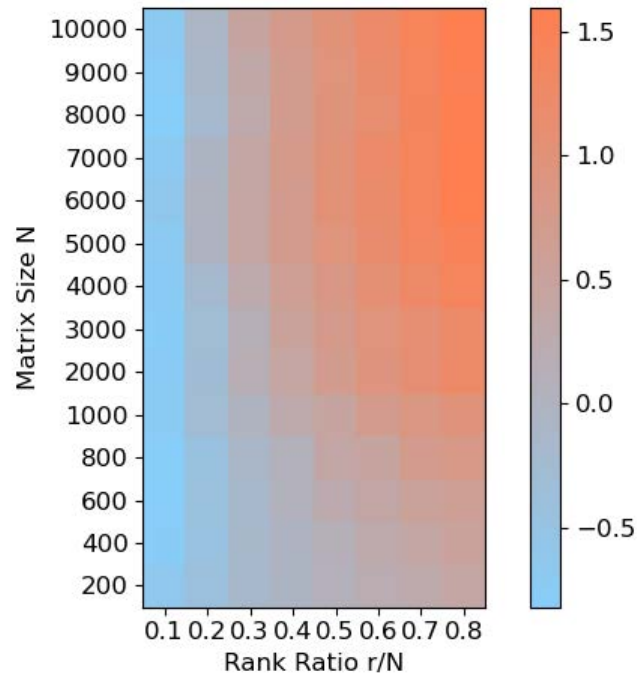}
         \caption{}
         \label{fig:heatmap}
     \end{subfigure}
     \caption{Performance characterization of LRSVD. (a) Execution time scaling versus matrix dimension $N$ (log-scale), showing superior NPU scalability over Intel MKL (CPU) and cuSOLVER ( GPU). \rev{The plotted rank ratio and algorithmic parameters are \rev{$r/N=0.1$, $n_{\text{loop}\_for\_QR} = 2$, $\text{tolerance}_{\text{QR}} = 0.005$}.} (b) Performance sensitivity heatmap across matrix size $N$ and rank ratio $r/N$,  \rev{where each grid cell represents the execution time ratio of low-rank solver executed on the NPU to H800 cuSOLVER}; blue regions identify the regimes where the NPU outpaces the \rev{H800} GPU.}
     \label{fig:svd_heatmap}
\end{figure}

\textbf{Analysis of $n_{loop}$:} The reported experiments use $n_{loop}=2$, matching the configuration in Fig.~\ref{fig:svd_heatmap}. This repeated projection suppresses the dominant single-precision orthogonality error without adding unnecessary passes.

\paragraph{Adaptive Convergence Thresholds}
Theoretical algorithms check if the norm $\|q_{k+1}\| = 0$ to determine subspace exhaustion. In floating-point arithmetic, strictly zero is rarely achieved. We introduce a relative error threshold $\epsilon$ tailored for the Ascend single-precision environment:
\begin{equation}
    \epsilon = 0.005
\end{equation}
This is the QR convergence tolerance used for the plotted results, together with $r/N=0.1$ and two QR refinement loops.

\paragraph{Validation}
To evaluate the resulting solver, we compare the Ascend implementation against three baselines: (i) full-rank SVD on an Intel Xeon Platinum 8168 CPU with MKL, (ii) an optimized CPU-based LRSVD, and (iii) a GPU-accelerated randomized SVD on NVIDIA \rev{H800/}A800 using cuSOLVER. Matrix sizes range from $N=1{,}000$ to $10{,}000$, covering representative single-node scientific problem sizes.

Fig.~\ref{fig:svd_heatmap}(a) shows that the Ascend 910B and 910C implementations consistently outperform all baselines and achieve especially strong gains at large matrix sizes. For $N=10{,}000$, the NPU solver delivers roughly an order-of-magnitude speedup over both the optimized CPU solver and the GPU cuSOLVER baseline. This improvement comes from offloading the dominant sketching and projection phases to the NPU while avoiding performance loss in the control-heavy stages. Fig.~\ref{fig:svd_heatmap}(b) further shows that the hybrid design is advantageous in the low-rank regime ($r/N<0.2$), the regime most relevant to model reduction and data compression. \rev{These results demonstrate that low-rank linear algebra workloads can be efficiently mapped to AI-oriented NPUs when execution stages are partitioned according to numerical sensitivity and hardware characteristics.}

\subsubsection{Precision-Emulated Matrix Multiplication (SGEMM-cube)}

SGEMM-cube is the third precision-gap application study. It combines numerical decomposition with a cache-aware execution pipeline to provide FP32-accurate dense matrix multiplication on FP16-native Cube units, complementing the solver-level strategies used by HPL-MxP and LRSVD.

\paragraph{Motivation and the Precision Gap Challenge}

General matrix multiplication (GEMM) is a foundational kernel in scientific computing, underpinning solvers for PDEs, quantum chemistry, and data-driven physics models. Its workload characteristics, high arithmetic intensity, regular data access, and structured reuse, make GEMM a prototypical compute-bound kernel on modern architectures. For numerical stability and reproducibility, scientific applications predominantly rely on FP32 or FP64 arithmetic.

However, emerging AI accelerators, including the Ascend NPU, are architected around Tensor Engines (Cube Units) that maximize throughput for low-precision formats such as FP16 and BF16. These formats are sufficient for deep learning but insufficient for many scientific workloads that demand FP32 fidelity. On Ascend 910A, FP32 operations are primarily executed on the Vector Units, whose peak throughput is two order of magnitude lower than that of the Cube Units. This architectural asymmetry creates a fundamental \emph{precision gap}: scientific workloads require FP32 accuracy, while the hardware’s highest-performance compute engines natively accept only lower-precision inputs.

SGEMM-cube is designed to bridge this gap. It is a mixed-precision dense linear algebra kernel that preserves FP32 semantics while offloading the bulk of computation to the high-throughput Cube Units. Rather than treating precision as a fixed hardware constraint, SGEMM-cube elevates it to an algorithmic design dimension. This shifts optimization from being purely compute-bound to being jointly algorithm- and architecture-driven, enabling AI-oriented tensor hardware to be systematically repurposed for high-performance scientific computing.

\paragraph{Mapping to Ascend and Precision Strategy}

\begin{figure}[t]
    \centering
    \includegraphics[width=\linewidth,trim=25 385 23 233, clip]{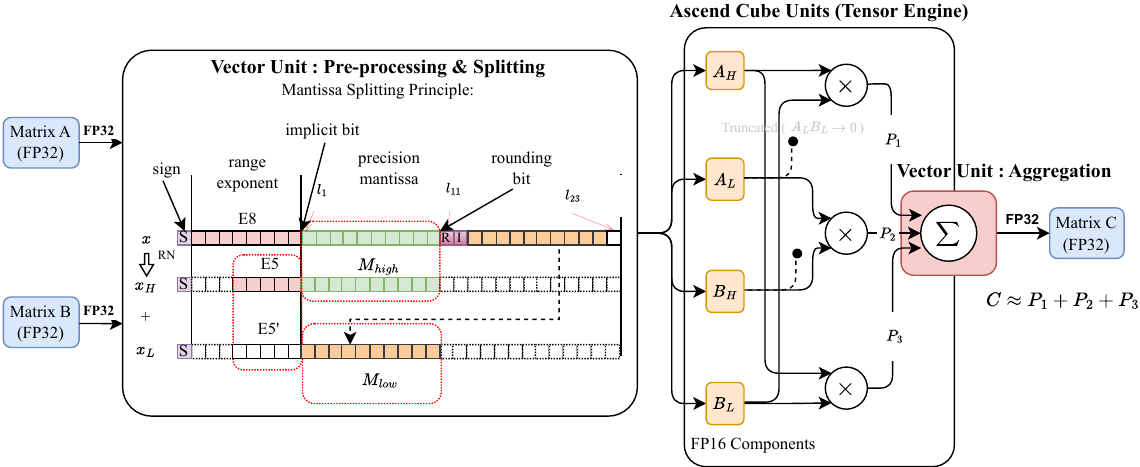}
    \caption{Overview of the SGEMM-cube algorithmic approach. FP32 inputs are split into FP16 High/Low pairs. Three distinct FP16 products are computed on the Cube Units and accumulated to recover FP32 precision.}
    \label{fig:sgemm_cube_workflow}
\end{figure}

To execute FP32 GEMM on BF16/FP16-native Cube Units, SGEMM-cube adopts a mantissa-splitting strategy that decomposes each FP32 operand into two lower-precision components. As illustrated in Fig.~\ref{fig:sgemm_cube_workflow}, an FP32 value $x$ is split into a high part $x_H$ and a low part $x_L$, such that $x \approx x_H + x_L$. The high part retains the exponent and the most significant mantissa bits, while the low part captures the residual. This decomposition preserves most of the numerical information while making the operands compatible with Cube Unit input formats.

For two FP32 matrices $A$ and $B$, the product can be expanded as:
\begin{equation}
\label{eq:sgemm_split}
\begin{aligned}
C &= (A_H + A_L)\,(B_H + B_L) \\
  &= A_H B_H 
   + A_H B_L 
   + A_L B_H 
   + \cancelto{\mathbf{0}}{A_L B_L}
\end{aligned}
\end{equation}

The $A_LB_L$ term is negligible due to the reduced magnitude of both operands and is therefore omitted. As a result, one logical FP32 GEMM is transformed into three physical FP16 GEMMs. Each term is computed on the Cube Units using FP16 inputs with FP32 accumulation, after which the partial products are aggregated to recover FP32 precision.

From a systems perspective, SGEMM-cube is implemented as a cache-aware, pipelined execution engine. The runtime orchestrates tiling, buffering, and precision control as a unified workflow: FP32 inputs are decomposed into FP16 components, tiles are staged into on-chip buffers, Cube Units execute the three GEMMs, and partial sums are accumulated locally in FP32 before being written back to global memory. Double buffering overlaps data movement with computation, keeping the Cube Units saturated. This design tightly couples algorithmic decomposition with Ascend NPU’s memory hierarchy and execution model, transforming GEMM from a single kernel invocation into a structured, high-throughput execution strategy.

\paragraph{Performance and Accuracy Evaluation}

We evaluate \texttt{sgemm-cube} on 910A, comparing it against three baselines: the native half-precision GEMM (\texttt{hgemm}) on 910A, the vendor-optimized single-precision GEMM from CANN library (\texttt{cann sgemm}) on 910B, and a standard CPU-based OpenBLAS implementation serving as the numerical gold standard.

\begin{figure}[t]
     \centering
     \begin{subfigure}[b]{0.48\columnwidth}
         \centering
         \includegraphics[width=\textwidth]{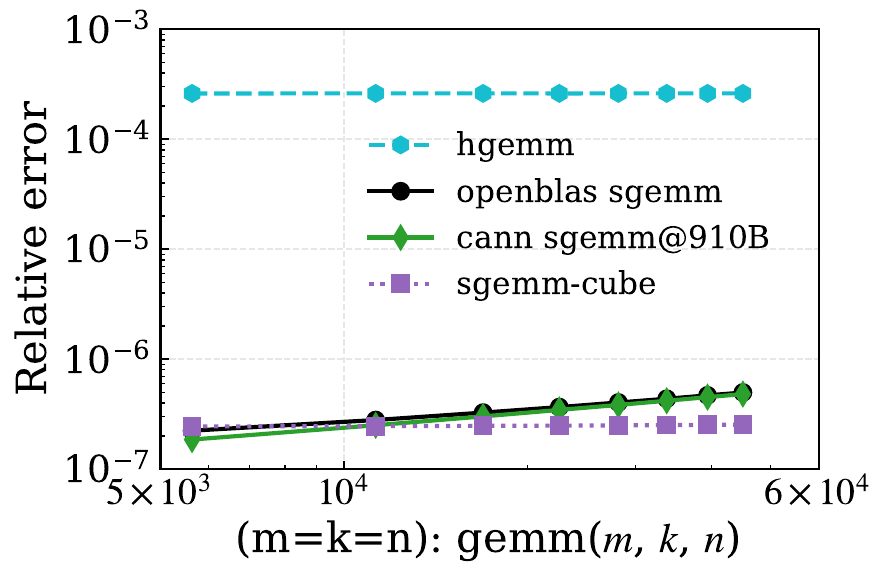} 
         \caption{}
         \label{fig:sgemm_cube_accuracy}
     \end{subfigure}
      \hfill
     \begin{subfigure}[b]{0.48\columnwidth}
         \centering
          \includegraphics[width=\textwidth] {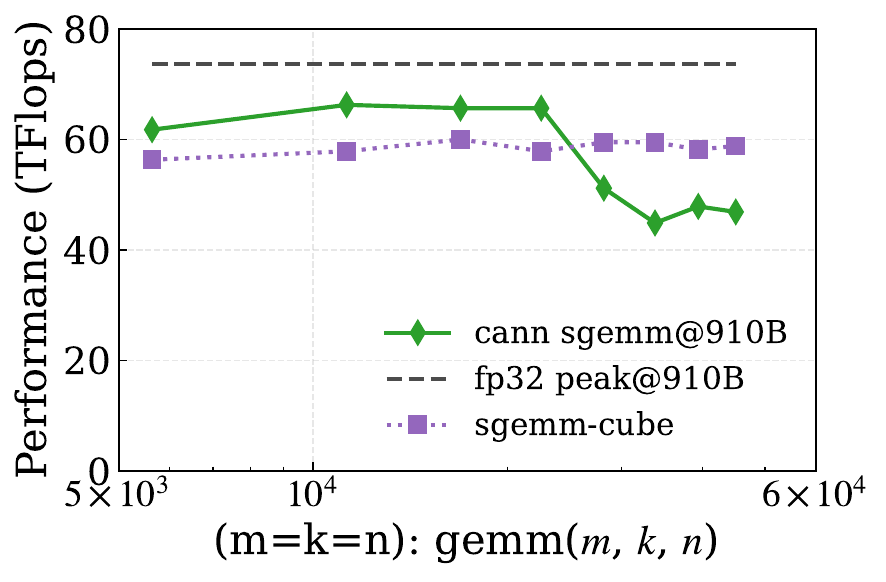}
         \caption{}
         \label{fig:sgemm_cube_perf}
     \end{subfigure}
     
     \caption{(a): Relative error across increasing matrix sizes. \texttt{sgemm-cube} achieves accuracy close to FP32 \texttt{sgemm} while significantly improving over native FP16 \texttt{hgemm}. (b): Performance (TFLOPS) comparison between CANN \texttt{sgemm} and \texttt{sgemm-cube}. 
    \texttt{sgemm-cube} sustains high throughput and avoids the performance drop observed in vendor \texttt{sgemm} at larger problem sizes, demonstrating both numerical robustness and stable scalability.}
     \label{fig:sgemm_cube_accuracy_perf}
\end{figure}

\textbf{Numerical Fidelity Verification.}
Fig.~\ref{fig:sgemm_cube_accuracy_perf} (a) quantifies the relative error against the OpenBLAS baseline. The results empirically validate the precision gap challenge: native \texttt{hgemm} yields errors in the order of $10^{-4}$, confirming its unsuitability for scientific kernels requiring rigorous convergence. In contrast, \texttt{sgemm-cube} effectively bridges this gap. By reconstructing the mantissa, it achieves an error magnitude of $10^{-7}$, matching the IEEE 754 single-precision standard. As matrix size increases, \texttt{cann sgemm} shows a slight drift in error, whereas \texttt{sgemm-cube} maintains a flat error profile. This confirms that the mantissa-splitting strategy successfully preserves FP32 semantics without accumulation divergence, even at large scales.

\textbf{Throughput and Scalability.} 
Fig.~\ref{fig:sgemm_cube_accuracy_perf} (b) presents the performance in TFLOPS. While the vendor implementation (\texttt{cann sgemm}) achieves higher peak throughput at smaller problem sizes, it suffers from significant performance volatility. Specifically, beyond $N=20,000$, the vendor kernel exhibits a sharp performance degradation (dropping from $\sim65$ to $\sim45$ TFLOPS), likely due to suboptimal cache tiling. Conversely, \texttt{sgemm-cube} exhibits excellent algorithmic scalability. It sustains a stable throughput of approximately $60$ TFLOPS, utilizing nearly $75\%$ of the theoretical peak, regardless of problem size. For large-scale matrices typical in HPC workloads, \texttt{sgemm-cube} outperforms the vendor implementation, proving that our cache-aware pipelined design provides the deterministic performance required for large-scale scientific simulations.

\subsection{Bridging the Data-Movement and Execution Gap}

\subsubsection{Bandwidth-Oriented Quantum Simulation (PQSim)}

\paragraph{Gap and application characteristics}
High-performance classical quantum simulation remains essential for validating quantum algorithms, benchmarking devices, and exploring quantum dynamics beyond current hardware limits~\cite{gill2020quantum}. Among quantum simulation approaches, state vector simulation is the most general, but it is also the most memory-intensive: an $n$ -qubit state requires a complex vector of size $2^n$, and each gate update induces structured but bandwidth-dominated accesses to this state vector~\cite{Nielsen2010}.

For Ascend NPUs, the challenge is therefore not arithmetic throughput alone. Each one- or two-qubit gate applies only a small local transform, but the dominant cost lies in moving amplitude between HBM and on-chip buffers. PQSim serves as a representative workload for the data-movement gap: it has regular fine-grained parallelism, but its performance is governed by bandwidth and locality rather than by tensor peak.

\paragraph{Solution: gate fusion and cache-oriented execution}
PQSim targets state-vector simulation by organizing gate updates as bandwidth-oriented streams. The quantum state resides in HBM as a $2^n$-dimensional complex vector, and each gate updates amplitude groups through small local transforms. These operations exhibit regular strided accesses, making them suitable for NPU execution only if data reuse is explicit.

\begin{figure}[t]
    \centering
    \includegraphics[width=\linewidth, trim=10 60 10 60, clip]{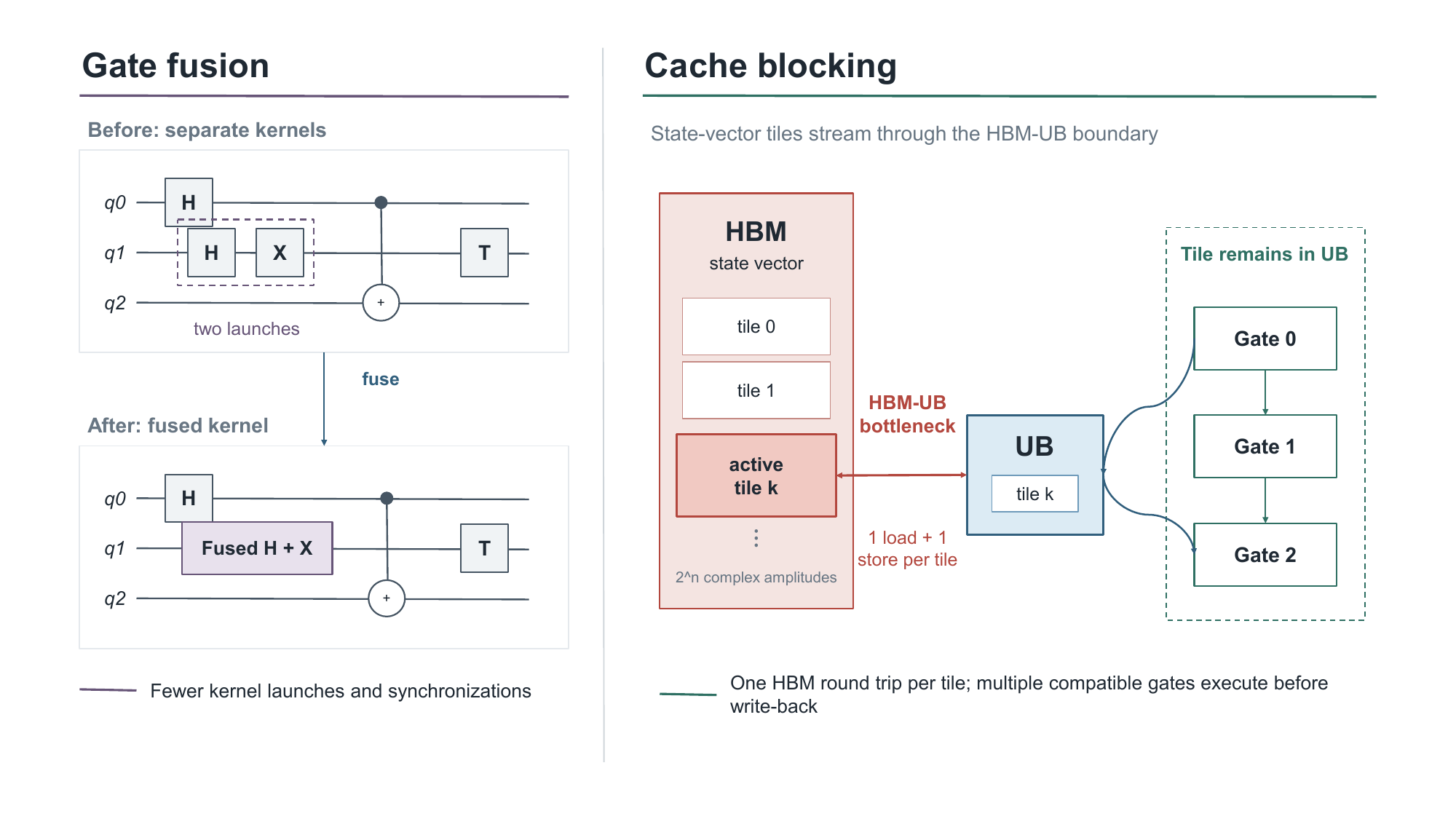}
    \caption{The systematic workflow of PQSim for high-performance quantum circuit simulation. The pipeline illustrates (1) input circuit pre-processing with gate fusion to minimize kernel overhead, 
    (2) a cache blocking strategy to optimize memory hierarchy utilization.}
    \label{fig:PQSim_workflow}
\end{figure}

To reduce launch overhead and synchronization frequency, PQSim applies circuit-level gate fusion (Fig.~\ref{fig:PQSim_workflow}). Consecutive gates operating on the same or nearby qubits are fused into a single kernel, so that multiple logical gate operations are executed within one invocation.

The second optimization targets the HBM--UB bottleneck directly. Because the full state vector is larger than the UB, PQSim partitions it into tiles and streams them through the UB. For gates acting on low-index qubits, all amplitude pairs involved in an update reside within a bounded tile. When several such gates appear consecutively, PQSim applies them back-to-back while the tile remains in UB, before writing it back to HBM. This transforms multiple HBM$\leftrightarrow$UB transfers into a single load--compute--store sequence, increasing on-chip reuse and reducing effective memory traffic.

\paragraph{Extended state-vector implementation detail}
Given our goal of providing a general-purpose, high-fidelity simulation platform, we adopt the state-vector approach in the current implementation.At its core, quantum computation corresponds to the unitary evolution of a closed quantum system. In the simulator, the state of an $n$-qubit register is represented by a state vector $\ket{\psi} \in \mathbb{C}^{2^n}$, whose components (amplitudes) are stored as single- or double-precision floating-point numbers. Each quantum gate implements a unitary transformation $U$ that updates this state via $\ket{\psi} \mapsto U\ket{\psi}$.

In the quantum circuit model , universal quantum computation can be achieved using only single- and two-qubit gates . Although the full unitary matrix for an $n$-qubit gate is formally a $2^n \times 2^n$ matrix, constructing it explicitly is both unnecessary and computationally prohibitive.Instead, we exploit the local structure of quantum gates. For a single-qubit gate $G$ acting on the $i$-th qubit, the effective transformation is applied through the operator  
$$
U = I(1) \otimes I(2) \otimes ... \otimes U(i) \otimes ... \otimes I(n),
$$  
where $I$ denotes the $2 \times 2$ identity matrix. Rather than forming $U$ explicitly, the state vector is updated by applying the $2 \times 2$ matrix $G$ to pairs of amplitudes that differ only in the $i$-th bit of their binary index (i.e., indices where the $i$-th qubit is 0 or 1). This reduces the operation to a series of independent $2$-dimensional transformations across the entire state vector.

Similarly, a two-qubit gate acting on qubits $i$ and $j$ (with $i < j$) corresponds to a $4 \times 4$ unitary matrix. The update is performed by grouping amplitudes into quadruplets whose indices share identical values on all qubits except $i$ and $j$; within each group, the four basis states $|00\rangle, |01\rangle, |10\rangle, |11\rangle$ (relative to qubits $i,j$) form a local subspace on which the $4 \times 4$ gate is applied directly.

This strategy avoids explicit construction of large matrices and instead performs structured, in-place updates of the state vector, achieving optimal computational efficiency while preserving numerical accuracy. Further implementation details, including memory layout and parallelization strategies, are discussed below.

\textbf{Gate Fusion Method} To reduce kernel launch overhead and communication frequency, PQSim performs circuit-level gate fusion (Fig.~\ref{fig:PQSim_workflow}). Consecutive gates operating on the same or nearby qubits are combined into a single composite kernel, so that multiple logical gate operations are executed within one NPU invocation. This reduces synchronization points in multi-node runs and significantly lowers the number of kernel dispatches on a single node~\cite{10.1145/3126908.3126947}.

\textbf{Cache Blocking.} To further mitigate the memory bottleneck between HBM and on-chip Unified Buffer (UB), PQSim employs a cache blocking technique tailored to the structure of quantum gate operations. 
During execution, the state vector is streamed from HBM into UB, updated in-place, and then written back to HBM (Fig.~\ref{fig:PQSim_workflow}). 
Since the full state vector is orders of magnitude larger than the UB capacity, it is partitioned into tiles of size \texttt{tileSize} and processed sequentially. For low-index (i.e., least-significant-qubit) gates, the distance between paired amplitudes involved in each update is bounded by \texttt{tileSize}. 
When a sequence of $N$ such low-index gates appears consecutively in the circuit, PQSim reorders data movement and computation to maximize data reuse: 
a tile of the state vector is first loaded into UB, all $N$ gates are applied successively while the data reside on-chip, and only then is the tile written back to HBM. 
This transforms $N$ separate load/store operations into a single HBM--UB transfer, significantly reducing memory traffic and improving effective bandwidth utilization.

\paragraph{\rev{Ascend-specific implementation}}
\rev{Gate fusion and memory-aware state-vector execution are established optimizations, including in cuQuantum~\cite{bayraktar2023cuquantumsdkhighperformancelibrary}. PQSim's contribution is to combine gate fusion with UB-resident blocking on Ascend, allowing several gate updates to reuse a tile before a single HBM write-back and thereby reducing repeated HBM--UB traffic.}

\begin{figure}[t]
    \centering
    \includegraphics[width=\linewidth]{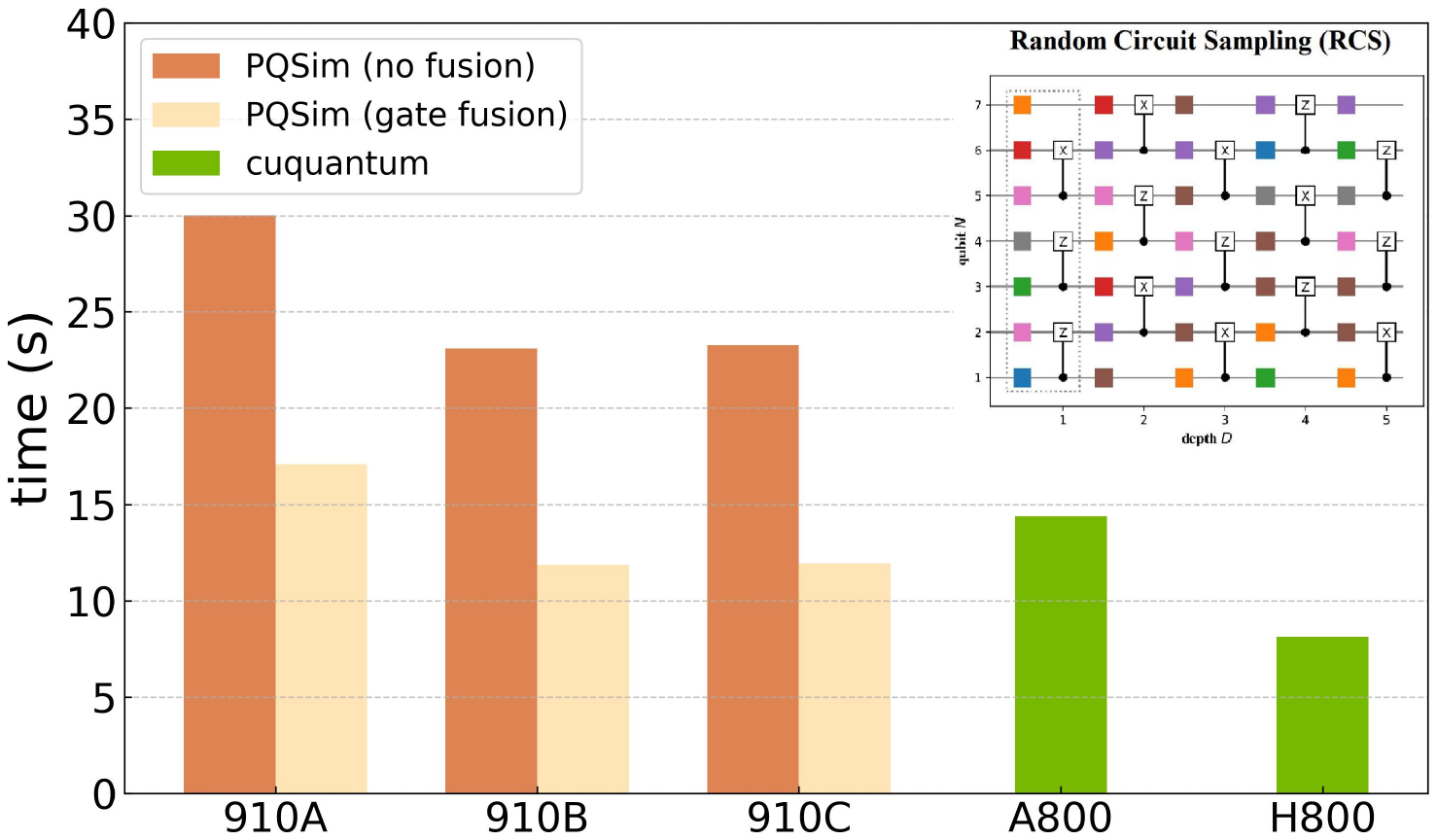}
    \caption{Performance benchmarking of PQSim on NPUs and GPUs using Random Circuit Sampling (RCS). The inset illustrates a typical 1D RCS circuit used for the benchmark.}
    \label{fig:PQSim_perf}
\end{figure}

\paragraph{Validation}
\rev{To isolate the impact of execution restructuring, we compare the unfused and fused PQSim implementations on the same Ascend platforms. This within-NPU comparison directly measures the benefit of gate fusion and memory-aware blocking independent of hardware generation.} We evaluate PQSim using the Random Circuit Sampling (RCS) benchmark~\cite{Google2019QuantumSupremacy} on Ascend 910A, 910B, and 910C, and compare against NVIDIA GPUs using cuQuantum 26.01.0. \rev{cuQuantum supports gate fusion for state-vector simulation~\cite{bayraktar2023cuquantumsdkhighperformancelibrary}.} As shown in Fig.~\ref{fig:PQSim_perf}, the simulation time for a 30-qubit, 30-layer circuit decreases consistently from 910A to 910C, reflecting the benefit of improved HBM bandwidth and architectural balance. On 910C, the baseline implementation without gate fusion completes in 22.2~s, compared with 27.5~s on 910A.

\rev{Within the Ascend implementations, gate fusion delivers an approximately $2\times$ speedup, supporting the conclusion that repeated state-vector movement is a dominant bottleneck.} With gate fusion enabled, PQSim on 910C (with bandwidth of 1.6TB/s) completes the benchmark in 11.4~s, outperforming NVIDIA A800 (with bandwidth of 2.0TB/s) with cuQuantum (14.3~s) by 25.4\%. \rev{Under this benchmark and software configuration, the result indicates that nominal HBM bandwidth alone does not determine performance; the platform-specific implementation of gate execution and memory reuse is also consequential.}

\subsubsection{Regularizing Irregular Monte Carlo for NPUs (SMC-X)}

\paragraph{Gap and application characteristics}
Monte Carlo (MC) is a method for equilibrium sampling in atomistic modeling for materials science and chemistry~\cite{LiuNPJ2025, doi:10.1021/acs.jctc.5c01614}. Unlike molecular dynamics, standard MC schemes such as Metropolis--Hastings are inherently sequential, which fundamentally limits their scalability on massively parallel architectures~\cite{PREIS20094468}. Existing parallel MC methods, including checkerboard and domain-decomposed approaches~\cite{PREIS20094468, 10.1145/3295500.3356149, ROMERO2020107473, mick2013gpu}, provide only limited generality, especially for machine-learning potentials (MLPs), where energy evaluations are irregular, data-intensive, and involve many-body interactions.

SMC-X (Scalable Monte Carlo at eXtreme) addresses this challenge through a dynamic link-cell decomposition and local interaction zones (LIZ), converting global MC updates into collections of local energy evaluations~\cite{LiuNPJ2025}. Even with this reformulation, the resulting workload remains difficult for AI-oriented NPUs: it exposes abundant fine-grained parallelism, but is dominated by irregular memory access, control divergence, and locality-sensitive data movement. SMC-X therefore serves as a representative test for the execution gap in its most difficult form: not just bandwidth-dominated computation, but irregular and divergence-prone scientific execution.

\paragraph{Extended SMC-X background}
Monte Carlo (MC) and molecular dynamics (MD) are the twin pillars of atomistic simulation, each with unique advantages in simulating equilibrium and non-equilibrium system, respectively. While MD naturally lends itself to parallelism, MC simulations have historically faced challenges due to the sequential nature of standard algorithms such as Metropolis-Hastings \cite{PREIS20094468}. Although simple models such as Ising \cite{PREIS20094468, 10.1145/3295500.3356149, ROMERO2020107473} and Lennard-Jones \cite{mick2013gpu} can exploit parallelism via checkerboard algorithms and their extensions, general short-range interactions, especially those involving machine learning potentials (MLP), lack scalable parallel MC methods. The SPMC method \cite{sadigh2012scalable}, implemented in the widely used LAMMPS package \cite{THOMPSON2022108171}, represents one of the few solutions that allows parallel MC simulations for arbitrary short-range interactions. However, SPMC’s parallelism is fundamentally constrained by its reliance on static domain decomposition, which requires large spatial domains and limits the scalability beyond one million atoms.

\begin{figure}[h]
  \centering
  \includegraphics[width=\linewidth]{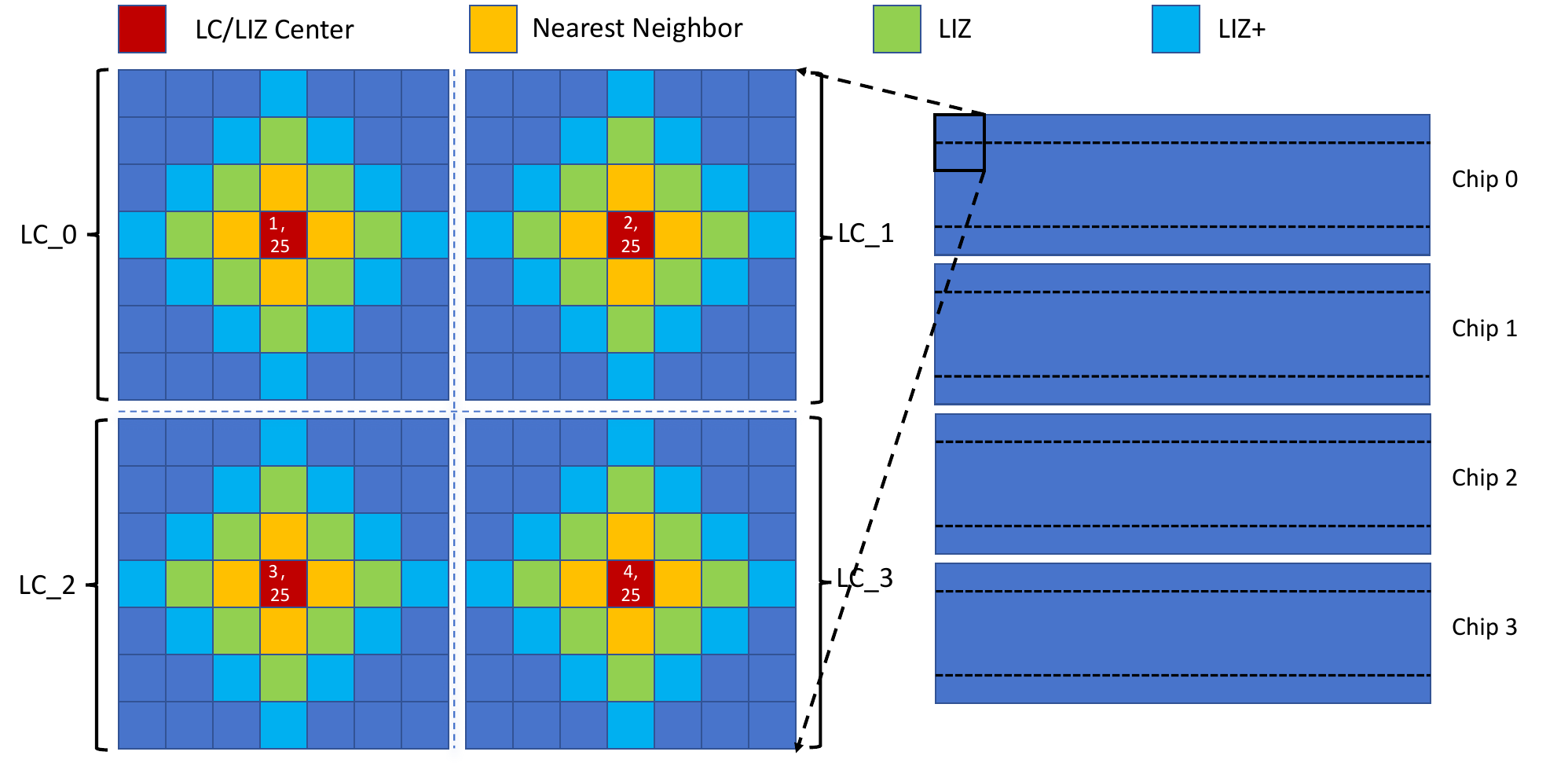}
  \caption{Schematics to illustrate the SMC-X algorithm. The dynamical LC is employed to isolate MC moves of lattice sites with different $i_C$ but the same $i_A$ index, the LIZ is used to evaluate the energy changes from MC trial, and the lattice is distributed over multiple chips. All other sites needed for evaluating the energy change is labeled as LIZ+.} \label{fig:SMC_algorithm}
\end{figure}

The recently introduced SMC-X (Scalable Monte Carlo at eXtreme) method \cite{LiuNPJ2025} is a fundamental advance for massively scalable Monte Carlo simulations of arbitrary short-range interactions. This breakthrough is enabled by two key innovations, as illustrated in Fig.~\ref{fig:SMC_algorithm}. First, the method employs a dynamic link-cell decomposition scheme. At each mini-step, the entire lattice is repartitioned into optimally sized link cells, ensuring the updated atom always resides at its cell's center. This not only guarantees the satisfaction of the detailed-balance condition, but also achieves substantially higher parallelism compared to SPMC \cite{sadigh2012scalable}.
Second, SMC-X introduces local interaction zones (LIZ) to efficiently handle non-pairwise interactions. Energy changes from Monte Carlo (MC) trials are computed through localized evaluations of post-trial energies, leveraging fine-grained parallelism to perform local energy calculations simultaneously for significant speedups. SMC-X successfully overcomes the long-standing challenge of parallel MC updates, and has demonstrated the capability to simulate one billion atoms on a single GPU \cite{LiuNPJ2025}.
This capability not only surpasses all previous MLP-based MC implementations, but also competes favorably with SOTA MLP-based MD. The extended technical report also records the related SMC-GPU result with a per-chip capacity of four billion atoms and scaling to 32 GPUs.

\begin{figure}[h]
  \centering
  \includegraphics[width=\linewidth]{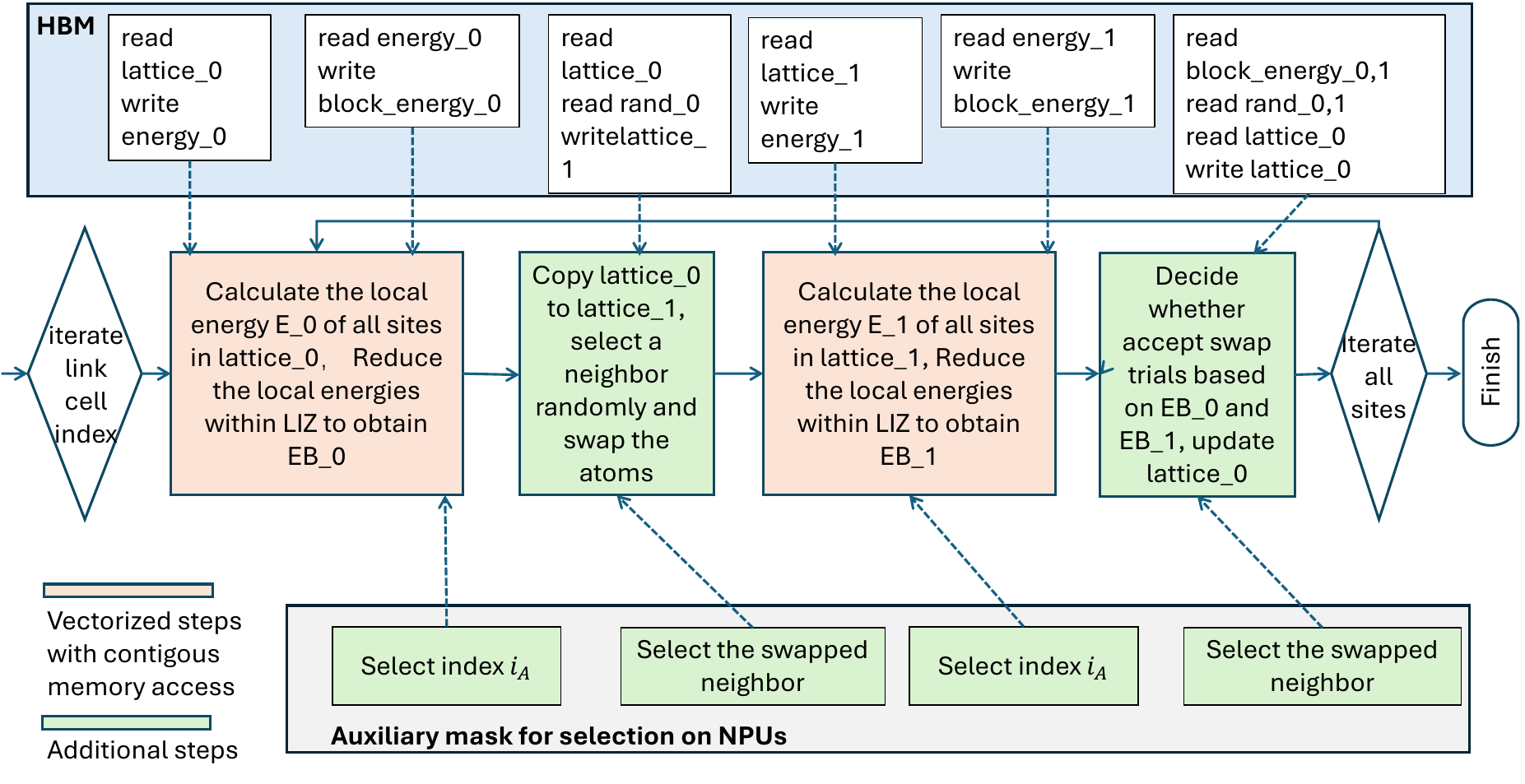}
  \caption{Architectural mapping and execution flow of the SMC-NPU algorithm. Green blocks highlight vector-optimized kernels adapted from SMC-X; orange blocks indicate NPU-specific adaptations for parallel execution.}
  \label{fig:NPU-flowchart}
\end{figure}

\paragraph{Solution: irregular-to-regular restructuring}
To make SMC-X efficient on Ascend, we redesign the update pipeline so that irregular MC behavior is transformed into vectorizable, memory-regular execution (Fig.~\ref{fig:NPU-flowchart}). The first step is to decouple local energy evaluation from Metropolis acceptance through a dual-lattice organization: the primary lattice stores atomic configurations, while an auxiliary lattice buffers post-trial local energies. This converts sequential accept/reject logic into batched operations, allowing all Metropolis criteria within a mini-step to be evaluated in parallel.

We then adapt the link-cell (LC) and local interaction zone (LIZ) mechanisms to the NPU execution model. Each LC/LIZ stage is implemented as a dedicated NPU kernel, and control-flow divergence is replaced with masked vector selection. This enables efficient SIMD execution across lattice sites while preserving the semantics of the original MC procedure.

Finally, memory latency is mitigated through explicit use of the Unified Buffer. Contiguous blocks of local energies and coordinates are staged in UB, and prefetching is used to overlap data movement with computation. Together, these changes transform SMC-X from a highly irregular update process into a batched, vector-oriented execution stream.

\paragraph{\rev{NPU-specific execution mapping}}
\rev{The contribution is the NPU-specific realization of SMC-X rather than the general concept of regularizing Monte Carlo execution. The implementation decouples energy evaluation from acceptance, replaces divergent branches with masked vector selection, and stages LC/LIZ data through UB. Together, these changes convert the existing local-update formulation into batched, locality-aware NPU kernels while preserving the Monte Carlo update semantics.}

\paragraph{Extended NPU implementation detail}
\textbf{Vectorization and memory access coalescing}
The original SMC-GPU algorithm \cite{LiuNPJ2025} involves a significant amount of irregular memory access. While such access patterns are efficiently handled by GPUs, they pose a major challenge on NPUs, which lack hardware support for flexible data movement. To address this limitation, the SMC-NPU algorithm introduces a dual-lattice structure, maintaining two separate lattices to store atomic configurations before and after swap trials. This design decouples the two critical computational phases, local energy calculation, and Metropolis updates. The local energy calculation steps, which dominate the computational workload, benefit from embarrassingly parallel execution and fully contiguous memory access. Although padding neighbor vectors to match the SIMD width of the NPU introduces some overhead, the contiguous memory access pattern is necessary to efficiently use the 2048-bit AIV. Note that since AIV does not support the INT8 datatype implemented for SMC-GPU, the atomic species are stored as INT16, which can compromise the memory usage and simulation throughput in SMC-NPU, as compared to SMC-GPU.

As previously mentioned, due to limitations in the NPU programming model, conditional branches and scalar operations must be re-implemented as masked vector operations. The vector processing unit provides a \texttt{select} operation, which enables element-wise selection between two vectors of the source operand. The result is written to a destination vector, and the operation is controlled by a mask, allowing conditional execution on a per-element basis:
\begin{equation}
    C = \text{Select(A, B, mask)}
\end{equation}
The mask perform the following:
\begin{itemize}
    \item When a mask bit is 1, the corresponding element is taken from the first source operand vector.
    \item When a mask bit is 0, the element is selected from the second source operand vector.
\end{itemize}
Note that some masks are generated by multiple steps with other auxiliary quantities. For instance, the neighbor-swap mask is obtained via bit-wise AND operations between a mask for swap-neighbor selection, and a mask for identifying $i_A$ index handled in each sequential mini-step.

\textbf{Hardware--software parallelism mapping}
The parallelism in SMC-X spans multiple algorithmic dimensions, including the energy model, LIZ, LC, and lattice. In contrast, the NPU computing platform offers hardware parallelism through vector/cube units, AI cores, dies/chips, and compute nodes. 
To achieve high performance, it is essential to carefully design a mapping strategy that aligns algorithmic and hardware parallelism across their respective hierarchical levels. Such a strategy must account for data locality, regular memory access patterns, and latency hiding to fully exploit the available hardware resources. While GPUs typically expose a more fine-grained programming model, owing to their use of 3D threads and blocks in the CUDA abstraction, the programming model of NPUs relies on explicitly managing vector/cube units and AI cores. As a result, the mapping strategy adopted in SMC-NPU differs significantly from that in SMC-GPU \cite{LiuNPJ2025}.

The most important difference is that the LIZ and LC degrees of parallelism are treated on the same footing in SMC-NPU, since both correspond to local energy calculations, which become embarrassingly parallel due to the use of two lattices in SMC-NPU. The total degree of parallelism available in these calculations scales linearly with the number of atoms in the lattice. At the NPU die level, this large degree of parallelism is exploited through both the architectural hierarchy and the fine-grained SIMD capability of the AIV units. Specifically, the effective hardware parallelism is given by: $N_{AIV} \times L_{AIV}/N_{bit}$, where $N_{AIV} = 48$ is the number of AIVs, $L_{AIV}$ is the SIMD width (in bits) of each AIV, and $N_{bit}=16$ denotes the bit-width of the data. In the mapping strategy, the innermost $z$-dimension of the lattice is assigned to the SIMD lanes within each AIV unit, while the $y$ and $x$ dimensions are distributed across the different AIVs. This layout ensures efficient vectorization and utilization of the NPU's hierarchical parallelism for local energy computations.

\begin{figure}[h]
  \centering
  \includegraphics[width=\linewidth]{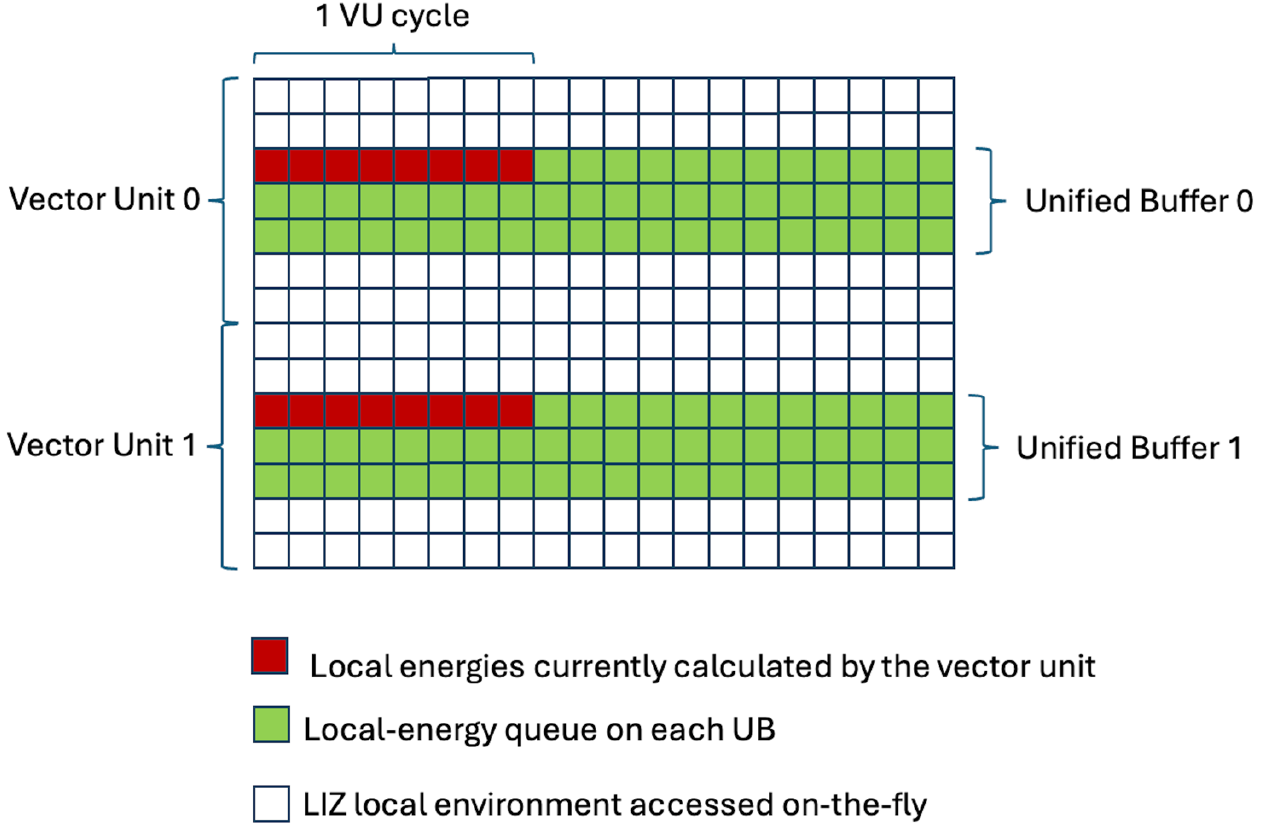}
  \caption{Schematic of local-energy vectorization in SMC-NPU. Queues of contiguous SIMD operands are staged in the Unified Buffer to hide memory latency.} \label{fig:cal_energy}
\end{figure}

\textbf{Memory Latency Hiding}
A critical distinction between GPU and NPU architectures lies in their cache hierarchy. As illustrated in Fig.~\ref{fig:NPU_arch}, the NPU Vector-Core path does not expose a GPU-like, automatically managed fine-grained L1 cache, which necessitates manual data movement optimization to hide HBM latency. This is particularly crucial for the local-energy kernel, which is the dominant computational hotspot. In practice, latency hiding is achieved by pre-fetching queued data into the unified buffer (UB) of the AI cores, as illustrated in Fig.~\ref{fig:cal_energy} based on the 2D square lattice case. For a 2048-bit AIV, 64 FP32 local energies can be computed simultaneously. To maximize buffer unit (BU) utilization and effectively hide memory latency, we carefully determine the number of local energies assigned to each VU, as demonstrated in Fig.~\ref{fig:cal_energy}.

It should be emphasized that the input features representing the local chemical environment information of each atom are computed on-the-fly and seamlessly integrated into both the local energy kernel and lattice updating kernel. This design creates a \textbf{tight coupling} between the machine learning potential (MLP) evaluation and atomistic simulation, which not only substantially reduce the HBM memory-footprint, but also significantly increase the efficiency by avoiding unnecessary data movement via better data locality.

\textbf{PBC ghost layer}
Another key difference between SMC-NPU and SMC-GPU lies in how periodic boundary conditions (PBC) are handled in the lattice. On NPUs, an atomic virtual layer is introduced to manage atomic exchanges across periodic boundaries, which is a standard technique for simulating crystalline systems. The reference GPU implementation does not require such a layer because it can efficiently determine whether an exchanged atom lies across the block boundary and directly access the corresponding data.
However, on Ascend processors, the SIMD (Single Instruction, Multiple Data) architecture makes such conditional checks inefficient or infeasible. As a result, the atomic virtual layer becomes necessary, even when the entire lattice supercell resides on a single accelerator. This virtual layer must be updated after each mini-step to maintain data consistency and ensure correct computation across periodic boundaries.

\paragraph{Validation}
We evaluate SMC-X on a representative chemically complex alloy system, $\rm{Fe_{29}Co_{29}Ni_{28}Al_7Ti_7}$, modeled by a quadratic short-range-order potential~\cite{LIU2023101018, MatterHEAQuantum}. We focus on the local energy kernel, which dominates MC runtime, and measure throughput in atom$\cdot$step/s to capture both compute efficiency and memory-system utilization.

Table~\ref{tab:opt} reports the per-chip throughput of SMC-X on four 910C NPUs and compares it with state-of-the-art atomistic simulation frameworks on GPUs and CPUs. Using four 910C NPUs, SMC-X achieves $1.81 \times 10^7$ atom$\cdot$steps/s per chip for billion-atom simulations, surpassing all non-SMC implementations and approaching the throughput of SMC-X on 16 H800 GPUs~\cite{doi:10.1021/acs.jctc.5c01614}. This indicates that, once appropriately restructured, NPUs can sustain competitive throughput even for complex Monte Carlo simulations.

\rev{Communication is small at the evaluated billion-atom scale because the surface-to-volume ratio is low. With approximately 1B atoms per NPU, ghost-layer exchange accounts for less than 2\% of runtime; this is why Table~\ref{tab:opt} focuses on local-energy throughput.} The bottom portion of Table~\ref{tab:opt} further highlights the importance of architecture-specific restructuring by serving as an incremental ablation of the NPU implementation, quantifying the contribution of vectorization, loop/data-movement restructuring, and low-level API optimization. Through successive optimizations in the \texttt{cal\_energy()} kernel, vectorization of Hamiltonian summation, loop and data-movement restructuring, and improved low-level API usage, the runtime is reduced from 83.4~s to 0.015~s per site, approaching the A800 implementation at 0.0105~s. The improvement is therefore not incremental but structural: performance is recovered only after the algorithm is reorganized into a SIMD-friendly and locality-aware form. These results confirm that irregular scientific workloads can run efficiently on AI-oriented NPUs, but only after the execution pattern itself has been regularized.

\begin{table}
  \centering
  \caption{Comparative performance evaluation of SMC-NPU against state-of-the-art ML-accelerated atomistic simulation frameworks across diverse hardware platforms. \rev{Throughput is reported in atom$\cdot$steps/s per chip.} The top section reports the per-chip throughput measured in atom$\cdot$step/s. The bottom section provides a detailed breakdown of the incremental performance gains achieved through specific optimizations applied to the NPU \texttt{cal\_energy} kernel, with A800 performance as a reference.}
  \label{tab:opt}
    \begin{tabular}{c|c}
\hline
\hline
\textbf{Applications} & \textbf{Throughput} \\
\hline
SNAP with 27,900 V100, 2021 \cite{10.1145/3458817.3487400} & $1.08 \times 10^{6}$
\\
DeePMD on 27,300 V100, 2022 \cite{10.1145/3503221.3508425} & $4.11 \times 10^{5}$
\\
Allegro on 5,120 A100, 2023 \cite{10.1145/3581784.3627041} & ${2.47 \times 10^{6}}$ \\
TensorMD on 32,000 SW-CPU, 2025 \cite{10880101} & $4.06 \times 10^{5}$ \\
SMC-GPU on 16 H800, 2025 \cite{doi:10.1021/acs.jctc.5c01614} & $3.62 \times 10^{7}$ \\
SMC-NPU on 4 NPU, 1B atoms (This work) & ${{1.81 \times 10^{7}}}$\\
\hline
\textbf{\texttt{cal\_energy()} optimizations on NPU} & \textbf{Time(s)} \\
\hline
Straight-forward implementation & 83.4463 \\
Vectorization of Hamiltonian summation & 0.2709 \\
Optimization of loop and data movement & 0.0327 \\
Improving low-level API usage & 0.0150\\
\textrm{A800 implementation of \texttt{cal\_energy()}} & 0.0105\\
\hline
    \end{tabular}
\end{table}

\subsection{Energy-to-Solution Analysis}

\begin{figure}[t]
\centering
\includegraphics[width=\linewidth]{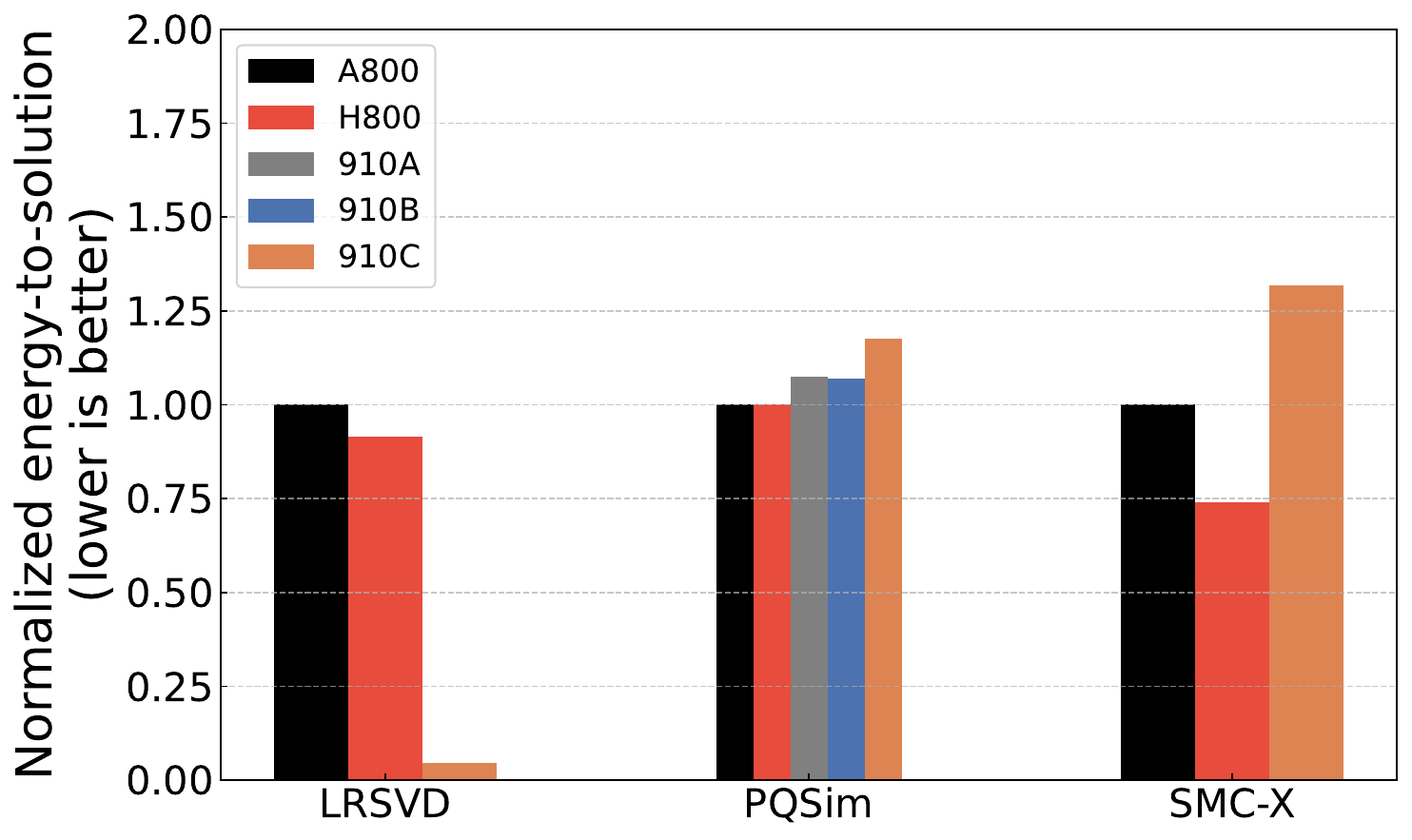}
\caption{\rev{Normalized energy-to-solution comparison of representative applications.
For each workload, the GPU baseline is normalized to 1.0.
The metric captures both execution time and average device power under the
evaluated implementations.}}
\label{fig:energy}
\end{figure}

\rev{
Beyond performance, energy consumption is an important consideration for
scientific computing on emerging AI-oriented accelerators. Figure~\ref{fig:energy}
reports normalized energy-to-solution for the subset of the five application
studies for which power measurements are available. For each workload, the corresponding
GPU baseline is normalized to 1.0; lower values indicate less energy required
to complete the same scientific task.
} \rev{
For LRSVD, the lower energy-to-solution on Ascend reflects not only the
accelerator characteristics but also the algorithmic difference between
the evaluated implementations. The Ascend implementation solves the
low-rank problem through a stage-specialized randomized LRSVD workflow,
whereas the GPU baseline uses cuSOLVER full-rank SVD.
}

\FloatBarrier
\section{Conclusion and Future Work}

This work studies scientific computing on AI-oriented NPUs through kernel characterization and cross-application evaluation. Using the Ascend 910 series as a representative tensor-centric architecture, we show that scientific workloads do not benefit uniformly from NPU peak throughput: Cube-friendly dense kernels scale well, whereas Vector-/memory-dominated kernels remain constrained by execution asymmetry and data movement. From this characterization, we identify three gaps that hinder direct deployment on AI-oriented NPUs: a precision gap for numerically sensitive algorithms, an execution mismatch for non-Cube-dominant workloads, and a data-movement gap across the memory and communication hierarchy. Across five application studies---HPL-MxP, LRSVD, SGEMM-cube, PQSim, and SMC-X---we show that these gaps can be bridged through heterogeneous stage partitioning, numerical stabilization, system-level mixed-precision mapping, FP32 precision emulation with cache-aware Cube execution, workload restructuring, and hierarchical data orchestration. Overall, these results show that AI-oriented NPUs can serve as effective scientific computing platforms, but only when numerical formulation, execution placement, and data movement are aligned with the architecture.

Some of the resulting insights are specific to the Ascend design, especially those tied to the exact Cube/Vector split, the UB/L0/L1 memory organization, and the CANN/HCCS software stack. However, several higher-level principles are broadly transferable to tensor-centric AI accelerators: heterogeneous stage partitioning for numerically sensitive workloads, restructuring irregular execution into accelerator-friendly forms, explicit orchestration of hierarchical data movement, and overlap of communication with computation. We therefore do not claim architecture-independent performance, but instead identify which optimization principles generalize and which remain platform-specific.

Several important limitations nevertheless remain. Our conclusions are drawn from one NPU family, and broader validation is needed to determine which optimization principles generalize across other AI-oriented processors. \rev{The mappings remain expert-driven, and broader standalone BLAS families beyond the SGEMM-cube case, sparse kernels, FFTs, and general stencil kernels are not covered; SMC-X provides only a stencil-like application case. On the evaluated NPU, sparse workloads are poorly matched to dense tensor datapaths without sparse-specific support, while non-emulatable FP64 operations still require CPU execution or dedicated high-precision mechanisms.} Future work will therefore focus on two directions: extending these methods to a wider class of tensor-centric architectures, and developing compiler- and runtime-assisted support for mixed precision, tiling, data layout, and communication scheduling to reduce manual tuning and improve portability.

\paragraph{Extended future directions.}
Looking ahead, the convergence of AI and scientific computing suggests that hardware architectures will continue to evolve towards extreme compute density with specialized memory hierarchies. Consequently, future scientific applications must embrace flexible precision paradigms and deep algorithmic adaptation to fully harness these emerging platforms:

\begin{itemize}
    \item \textbf{Flexible Precision Exploration:} To exploit the peak throughput of next-generation compute units (e.g., advanced Cube Units or Tensor Cores), solvers should explore beyond standard mixed-precision (FP32/FP64). Future work involves investigating finer-grained numerical formats (such as BF16, FP8, or adaptive precision) that can trade manageable accuracy loss for massive throughput gains while maintaining convergence guarantees.
    
    \item \textbf{Hardware-Aware Algorithmic Adjustment:} As the disparity between logic speed and memory bandwidth widens, algorithms must be redesigned to be intrinsically "cache-aware." This entails dynamically adjusting data layouts and tiling strategies to align with complex on-chip memory hierarchies (L1/L2 buffers), ensuring that high-performance compute units are constantly fed with data. We envision a shift towards algorithm-architecture co-design, where numerical methods are reshaped to fit the unique data movement patterns of AI-centric hardware.
\end{itemize}

\section*{Acknowledgments}
This work was supported in part by Guangdong S\&T Programme under Grant 2024B0101010003. We gratefully acknowledge the Bisheng C++ team for their sustained technical collaboration and for developing a robust, flexible, and high-performance programming framework for Ascend processors, which was essential to this study.

\bibliographystyle{unsrt}  
\bibliography{AscendSci}  

@article{
doi:10.1126/science.adu0801,
author = {Ewa Deelman  and Jack Dongarra  and Bruce Hendrickson  and Amanda Randles  and Daniel Reed  and Edward Seidel  and Katherine Yelick },
title = {High-performance computing at a crossroads},
journal = {Science},
volume = {387},
number = {6736},
pages = {829-831},
year = {2025},
doi = {10.1126/science.adu0801},
URL = {https://www.science.org/doi/abs/10.1126/science.adu0801},
eprint = {https://www.science.org/doi/pdf/10.1126/science.adu0801}}

@ARTICLE{9351692,
  author={Norrie, Thomas and Patil, Nishant and Yoon, Doe Hyun and Kurian, George and Li, Sheng and Laudon, James and Young, Cliff and Jouppi, Norman and Patterson, David},
  journal={IEEE Micro}, 
  title={The Design Process for Google's Training Chips: TPUv2 and TPUv3}, 
  year={2021},
  volume={41},
  number={2},
  pages={56-63},
  url = {https://doi.org/10.1109/MM.2021.3058217},
  doi={10.1109/MM.2021.3058217}}

@misc{zuo2025servinglargelanguagemodels,
      title={Serving Large Language Models on Huawei CloudMatrix384}, 
      author={Pengfei Zuo and Huimin Lin and Junbo Deng and Nan Zou and Xingkun Yang and Yingyu Diao and Weifeng Gao and Ke Xu and Zhangyu Chen and Shirui Lu and Zhao Qiu and Peiyang Li and Xianyu Chang and Zhengzhong Yu and Fangzheng Miao and Jia Zheng and Ying Li and Yuan Feng and Bei Wang and Zaijian Zong and Mosong Zhou and Wenli Zhou and Houjiang Chen and Xingyu Liao and Yipeng Li and Wenxiao Zhang and Ping Zhu and Yinggang Wang and Chuanjie Xiao and Depeng Liang and Dong Cao and Juncheng Liu and Yongqiang Yang and Xiaolong Bai and Yi Li and Huaguo Xie and Huatao Wu and Zhibin Yu and Lv Chen and Hu Liu and Yujun Ding and Haipei Zhu and Jing Xia and Yi Xiong and Zhou Yu and Heng Liao},
      year={2025},
      eprint={2506.12708},
      archivePrefix={arXiv},
      primaryClass={cs.DC},
      url={https://arxiv.org/abs/2506.12708v1}, 
}

@inproceedings{10.1145/3698038.3698535,
author = {Fu, Xinwei and Zhang, Zhen and Fan, Haozheng and Huang, Guangtai and El-Shabani, Mohammad and Huang, Randy and Solanki, Rahul and Wu, Fei and Diamant, Ron and Wang, Yida},
title = {Distributed Training of Large Language Models on AWS Trainium},
year = {2024},
isbn = {9798400712869},
publisher = {Association for Computing Machinery},
address = {New York, NY, USA},
url = {https://doi.org/10.1145/3698038.3698535},
doi = {10.1145/3698038.3698535},
booktitle = {Proceedings of the 2024 ACM Symposium on Cloud Computing},
pages = {961--976},
numpages = {16},
location = {Redmond, WA, USA},
series = {SoCC '24}
}

@article{IntelSaphaireRapids,
	author = {Siegmann, Eva and Harrison, Robert J. and Carlson, David and Chheda, Smeet and Curtis, Anthony and Coskun, Firat and Gonzalez, Raul and Wood, Daniel and Simakov, Nikolay A.},
	date = {2024/06/07},
	doi = {10.1007/s42979-024-02958-3},
	id = {Siegmann2024},
	isbn = {2661-8907},
	journal = {SN Computer Science},
	number = {5},
	pages = {623},
	title = {First Impressions of the Sapphire Rapids Processor with HBM for Scientific Workloads},
	url = {https://doi.org/10.1007/s42979-024-02958-3},
	volume = {5},
	year = {2024}}

@ARTICLE{10070122,
  author={Choquette, Jack},
  journal={IEEE Micro}, 
  title={NVIDIA Hopper H100 GPU: Scaling Performance}, 
  year={2023},
  volume={43},
  number={3},
  pages={9-17},
  url = {https://doi.org/10.1109/MM.2023.3256796},
  doi={10.1109/MM.2023.3256796}}

@article{JackNaturePhysics,
	author = {Dongarra, Jack and Keyes, David},
	date = {2024/10/01},
	doi = {10.1038/s42254-024-00750-z},
	id = {Dongarra2024},
	isbn = {2522-5820},
	journal = {Nature Reviews Physics},
	number = {10},
	pages = {621--627},
	title = {The co-evolution of computational physics and high-performance computing},
	url = {https://doi.org/10.1038/s42254-024-00750-z},
	volume = {6},
	year = {2024}}

@article{LiuNPJ2025,
	author = {Liu, Xianglin and Yang, Kai and Liu, Yongxiang and Zhou, Fanli and Fan, Dengdong and Pei, Zongrui and Xu, Pengxiang and Tian, Yonghong},
	date = {2025/08/20},
	doi = {10.1038/s41524-025-01762-8},
	id = {Liu2025},
	isbn = {2057-3960},
	journal = {npj Computational Materials},
	number = {1},
	pages = {267},
	title = {Revealing nanostructures in high-entropy alloys via machine-learning accelerated scalable Monte Carlo simulation},
	url = {https://doi.org/10.1038/s41524-025-01762-8},
	volume = {11},
	year = {2025}}

@article{sadigh2012scalable,
  title     = {Scalable parallel Monte Carlo algorithm for atomistic simulations of precipitation in alloys},
  author    = {Sadigh, Babak and Erhart, Paul and Stukowski, Alexander and Caro, Alfredo and Martinez, Enrique and Zepeda-Ruiz, Luis},
  journal   = {Physical Review B},
  volume    = {85},
  number    = {18},
  pages     = {184203},
  year      = {2012},
  month     = {May},
  doi       = {10.1103/PhysRevB.85.184203},
  url       = {https://doi.org/10.1103/PhysRevB.85.184203}
}

@article{mick2013gpu,
  title     = {GPU-accelerated Gibbs ensemble Monte Carlo simulations of Lennard-Jonesium},
  author    = {Mick, Jason and Hailat, Eyad and Russo, Vincent and Rushaidat, Kamel and Schwiebert, Loren and Potoff, Jeffrey},
  journal   = {Computer Physics Communications},
  volume    = {184},
  number    = {12},
  pages     = {2662--2669},
  year      = {2013},
  doi       = {10.1016/j.cpc.2013.06.020},
  url       = {https://doi.org/10.1016/j.cpc.2013.06.020}
}

@article{PREIS20094468,
title = {GPU accelerated Monte Carlo simulation of the 2D and 3D Ising model},
journal = {Journal of Computational Physics},
volume = {228},
number = {12},
pages = {4468-4477},
year = {2009},
issn = {0021-9991},
doi = {https://doi.org/10.1016/j.jcp.2009.03.018},
url = {https://www.sciencedirect.com/science/article/pii/S0021999109001387},
author = {Tobias Preis and Peter Virnau and Wolfgang Paul and Johannes J. Schneider}
}

@article{THOMPSON2022108171,
title = {LAMMPS - a flexible simulation tool for particle-based materials modeling at the atomic, meso, and continuum scales},
journal = {Computer Physics Communications},
volume = {271},
pages = {108171},
year = {2022},
issn = {0010-4655},
doi = {https://doi.org/10.1016/j.cpc.2021.108171},
url = {https://www.sciencedirect.com/science/article/pii/S0010465521002836},
author = {Aidan P. Thompson and H. Metin Aktulga and Richard Berger and Dan S. Bolintineanu and W. Michael Brown and Paul S. Crozier and Pieter J. {in 't Veld} and Axel Kohlmeyer and Stan G. Moore and Trung Dac Nguyen and Ray Shan and Mark J. Stevens and Julien Tranchida and Christian Trott and Steven J. Plimpton}
}

@article{ROMERO2020107473,
title = {High performance implementations of the 2D Ising model on GPUs},
journal = {Computer Physics Communications},
volume = {256},
pages = {107473},
year = {2020},
issn = {0010-4655},
doi = {https://doi.org/10.1016/j.cpc.2020.107473},
url = {https://www.sciencedirect.com/science/article/pii/S0010465520302228},
author = {Joshua Romero and Mauro Bisson and Massimiliano Fatica and Massimo Bernaschi}
}

@article{lin2024fastattention,
  title={Fastattention: Extend flashattention2 to npus and low-resource gpus},
  author={Lin, Haoran and Yu, Xianzhi and Zhao, Kang and Hou, Lu and Zhan, Zongyuan and Kamenev, Stanislav and Bao, Han and Hu, Ting and Wang, Mingkai and Chang, Qixin and others},
  journal={arXiv preprint arXiv:2410.16663},
  url={https://arxiv.org/abs/2410.16663},
  year={2024}
}

@inproceedings{yang2019high,
  author = {Yang, Kun and Chen, Yi-Fan and Roumpos, Georgios and Colby, Chris and Anderson, John},
title = {High Performance Monte Carlo Simulation of Ising Model on TPU Clusters},
year = {2019},
isbn = {9781450362290},
publisher = {Association for Computing Machinery},
address = {New York, NY, USA},
url = {https://doi.org/10.1145/3295500.3356149},
doi = {10.1145/3295500.3356149},
booktitle = {Proceedings of the International Conference for High Performance Computing, Networking, Storage and Analysis},
articleno = {83},
numpages = {15},
location = {Denver, Colorado},
series = {SC '19}
}

@article{pederson2023large,
  author    = {Pederson, Ryan and Kozlowski, John and Cho, Minkyu and Malone, Fionn D. and Lee, Joonho and Lewis, Adam G. M. and Beall, Jackson and Ganahl, Martin and Hauru, Markus and Blum, Volker and Vidal, Guifr{\'e}},
  title     = {Large-Scale Quantum Chemistry with Tensor Processing Units},
  journal   = {Journal of Chemical Theory and Computation},
  volume    = {19},
  number    = {1},
  pages     = {25--32},
  year      = {2023},
  publisher = {American Chemical Society},
  url       = {https://doi.org/10.1021/acs.jctc.2c00876},
  doi       = {10.1021/acs.jctc.2c00876}
}

@article{lewis2022large,
  author    = {Lewis, Adam G. M. and Beall, Jackson and Ganahl, Martin and Hauru, Markus and Mallick, Shrestha Basu and Vidal, Guifre},
  title     = {Large-scale distributed linear algebra with tensor processing units},
  journal   = {Proceedings of the National Academy of Sciences},
  volume    = {119},
  number    = {33},
  pages     = {e2122762119},
  year      = {2022},
  publisher = {National Academy of Sciences},
  url       = {https://doi.org/10.1073/pnas.2122762119},
  doi       = {10.1073/pnas.2122762119}
}

@article{trifan2022intelligent,
  author    = {Trifan, Anda and Gorgun, Defne and Li, Zongyi and Brace, Alexander and Zvyagin, Maxim and Ma, Heng and Clyde, Austin and Clark, David and Salim, Michael and Hardy, David J. and others},
  title     = {Intelligent resolution: Integrating Cryo-EM with AI-driven multi-resolution simulations to observe the SARS-CoV-2 replication-transcription machinery in action},
  journal   = {The International Journal of High Performance Computing Applications},
  volume    = {36},
  number    = {5-6},
  pages     = {603--623},
  year      = {2022},
  publisher = {SAGE Publications},
  url       = {https://doi.org/10.1177/10943420221113513},
  doi       = {10.1177/10943420221113513}
}

@INPROCEEDINGS{xue_unlocking,
  author={Xue, Weicheng and Yang, Kai and Liu, Yongxiang and Fan, Dengdong and Xu, Pengxiang and Tian, Yonghong},
  booktitle={SC24: International Conference for High Performance Computing, Networking, Storage and Analysis}, 
  title={Unlocking High Performance with Low-Bit NPUs and CPUs for Highly Optimized HPL-MxP on Cloud Brain II}, 
  year={2024},
  volume={},
  number={},
  pages={1-16},
  url={https://doi.org/10.1109/SC41406.2024.00088},
  doi={10.1109/SC41406.2024.00088}}

@article{xu2023fast,
  title={Fast algorithms for singular value decomposition and the inverse of nearly low-rank matrices},
  author={Xu, Chen and Xu, Weiwei and Jing, Kaili},
  journal={National Science Review},
  volume={10},
  number={6},
  pages={nwad083},
  year={2023},
  doi={10.1093/nsr/nwad083},
  url={https://doi.org/10.1093/nsr/nwad083},
  publisher={Oxford University Press}}

@article{halko2011finding,
  title={Finding structure with randomness: Probabilistic algorithms for constructing approximate matrix decompositions},
  author={Halko, Nathan and Martinsson, Per-Gunnar and Tropp, Joel A},
  journal={SIAM review},
  volume={53},
  number={2},
  pages={217--288},
  year={2011},
  doi={10.1137/090771806},
  url={https://doi.org/10.1137/090771806},
  publisher={SIAM}
}

@article{musco2015randomized,
  title={Randomized block krylov methods for stronger and faster approximate singular value decomposition},
  author={Musco, Cameron and Musco, Christopher},
  journal={Advances in neural information processing systems},
  volume={28},
  url={https://dl.acm.org/doi/10.5555/2969239.2969395},
  year={2015}
}

@article{tropp2019streaming,
  title={Streaming low-rank matrix approximation with an application to scientific simulation},
  author={Tropp, Joel A and Yurtsever, Alp and Udell, Madeleine and Cevher, Volkan},
  journal={SIAM Journal on Scientific Computing},
  volume={41},
  number={4},
  pages={A2430--A2463},
  year={2019},
  doi={10.1137/18m1201068},
  url={https://doi.org/10.1137/18M1201068},
  publisher={SIAM}
}

@inproceedings{10.1145/3773656.3773670,
author = {Schwarz, Angelika and Anders, Anton and Brower, Cole and Bayraktar, Harun and Gunnels, John and Clark, Kate and Xu, RuQing G. and Rodriguez, Samuel and Cayrols, Sebastien and Tabaszewski, Pawel and Podlozhnyuk, Victor},
title = {Guaranteed DGEMM Accuracy While Using Reduced Precision Tensor Cores Through Extensions of the Ozaki Scheme},
year = {2026},
isbn = {9798400720673},
publisher = {Association for Computing Machinery},
address = {New York, NY, USA},
url = {https://doi.org/10.1145/3773656.3773670},
doi = {10.1145/3773656.3773670},
booktitle = {Proceedings of the Supercomputing Asia and International Conference on High Performance Computing in Asia Pacific Region},
pages = {91--101},
numpages = {11},
location = {
},
series = {SCA/HPCAsia '26}
}

@article{10.1177/10943420251382476,
author = {Dongarra, Jack and Luszczek, Piotr},
title = {HPL-MxP benchmark: Mixed-precision algorithms, iterative refinement, and scalable data generation},
year = {2026},
issue_date = {Jan 2026},
publisher = {Sage Publications, Inc.},
address = {USA},
volume = {40},
number = {1},
issn = {1094-3420},
url = {https://doi.org/10.1177/10943420251382476},
doi = {10.1177/10943420251382476},
journal = {Int. J. High Perform. Comput. Appl.},
month = jan,
pages = {52--62},
numpages = {11}
}

@article{gill2020quantum,
  title={Quantum computing: A taxonomy, systematic review and future directions},
  author={Gill, Sukhpal Singh and Tuli, Shreshth and Xu, Muhan and Singh, Inderveer Chana and Dustdar, Schahram and Buyya, Rajkumar},
  journal={Software: Practice and Experience},
  year={2022},
  volume={52},
  number={1},
  pages={92--136},
  publisher={Wiley},
  url = {https://onlinelibrary.wiley.com/doi/abs/10.1002/spe.3039}
}

@Book{Nielsen2010,
  author    = {M. Nielsen and I. Chuang},
  publisher = {Cambridge University Press},
  title     = {Quantum Computation and Quantum Information},
  year      = {2010},
  address   = {Cambridge},
  edition   = {10th Anniversary Edition},
  isbn      = {9780511976667},
  month     = jun,
  doi       = {10.1017/cbo9780511976667},
  url       = {https://doi.org/10.1017/cbo9780511976667}
}

@inproceedings{10.1145/3126908.3126947,
author = {H\"{a}ner, Thomas and Steiger, Damian S.},
title = {0.5 petabyte simulation of a 45-qubit quantum circuit},
year = {2017},
isbn = {9781450351140},
publisher = {Association for Computing Machinery},
address = {New York, NY, USA},
url = {https://doi.org/10.1145/3126908.3126947},
doi = {10.1145/3126908.3126947},
booktitle = {Proceedings of the International Conference for High Performance Computing, Networking, Storage and Analysis},
articleno = {33},
numpages = {10},
location = {Denver, Colorado},
series = {SC '17}
}

@Article{Google2019QuantumSupremacy,
  author    = {Arute, Frank and Arya, Kunal and Babbush, Ryan and Bacon, Dave and Bardin, Joseph C. and Barends, Rami and Biswas, Rupak and Boixo, Sergio and Brandao, Fernando G. S. L. and Buell, David A. and Burkett, Brian and Chen, Yu and Chen, Zijun and Chiaro, Ben and Collins, Roberto and Courtney, William and Dunsworth, Andrew and Farhi, Edward and Foxen, Brooks and Fowler, Austin and Gidney, Craig and Giustina, Marissa and Graff, Rob and Guerin, Keith and Habegger, Steve and Harrigan, Matthew P. and Hartmann, Michael J. and Ho, Alan and Hoffmann, Markus and Huang, Trent and Humble, Travis S. and Isakov, Sergei V. and Jeffrey, Evan and Jiang, Zhang and Kafri, Dvir and Kechedzhi, Kostyantyn and Kelly, Julian and Klimov, Paul V. and Knysh, Sergey and Korotkov, Alexander and Kostritsa, Fedor and Landhuis, David and Lindmark, Mike and Lucero, Erik and Lyakh, Dmitry and Mandr{\`a}, Salvatore and McClean, Jarrod R. and McEwen, Matthew and Megrant, Anthony and Mi, Xiao and Michielsen, Kristel and Mohseni, Masoud and Mutus, Josh and Naaman, Ofer and Neeley, Matthew and Neill, Charles and Niu, Murphy Yuezhen and Ostby, Eric and Petukhov, Andre and Platt, John C. and Quintana, Chris and Rieffel, Eleanor G. and Roushan, Pedram and Rubin, Nicholas C. and Sank, Daniel and Satzinger, Kevin J. and Smelyanskiy, Vadim and Sung, Kevin J. and Trevithick, Matthew D. and Vainsencher, Amit and Villalonga, Benjamin and White, Theodore and Yao, Z. Jamie and Yeh, Ping and Zalcman, Adam and Neven, Hartmut and Martinis, John M.},
  title     = {Quantum supremacy using a programmable superconducting processor},
  journal   = {Nature},
  year      = {2019},
  volume    = {574},
  pages     = {505--510},
  issn      = {1476-4687},
  doi       = {10.1038/s41586-019-1666-5},
  url       = {https://doi.org/10.1038/s41586-019-1666-5},
}

@article{doi:10.1021/acs.jctc.5c01614,
	author = {Liu, Xianglin and Yang, Kai and Zhou, Fanli and Xu, Pengxiang},
	date = {2025/12/23},
	doi = {10.1021/acs.jctc.5c01614},
	isbn = {1549-9618},
	journal = {Journal of Chemical Theory and Computation},
	journal1 = {Journal of Chemical Theory and Computation},
	journal2 = {J. Chem. Theory Comput.},
	month = {12},
	number = {24},
	pages = {12784--12795},
	publisher = {American Chemical Society},
	title = {SMC-X: A Distributed, Scalable Monte Carlo Simulation Method for Chemically Complex Alloys},
	type = {doi: 10.1021/acs.jctc.5c01614},
	url = {https://doi.org/10.1021/acs.jctc.5c01614},
	volume = {21},
	year = {2025},
	year1 = {2025}}

@article{LIU2023101018,
title = {Machine learning for high-entropy alloys: Progress, challenges and opportunities},
journal = {Progress in Materials Science},
volume = {131},
pages = {101018},
year = {2023},
issn = {0079-6425},
doi = {https://doi.org/10.1016/j.pmatsci.2022.101018},
url = {https://www.sciencedirect.com/science/article/pii/S0079642522000998},
author = {Xianglin Liu and Jiaxin Zhang and Zongrui Pei}
}

@article{MatterHEAQuantum,
	author = {Pei, Zongrui and Gong, Yilun and Liu, Xianglin and Yin, Junqi},
	date = {2024/10/02},
	doi = {10.1016/j.matt.2024.05.035},
	isbn = {2590-2393},
	journal = {Matter},
	journal1 = {Matter},
	month = {2026/02/01},
	number = {10},
	pages = {3433--3446},
	publisher = {Elsevier},
	title = {Designing complex concentrated alloys with quantum machine learning and language modeling},
	type = {doi: 10.1016/j.matt.2024.05.035},
	url = {https://doi.org/10.1016/j.matt.2024.05.035},
	volume = {7},
	year = {2024},
	year1 = {2024}}

@ARTICLE{10880101,
  author={Wang, Xun and Meng, Xiangyu and Guo, Zhuoqiang and Li, Mingzhen and Liu, Lijun and Li, Mingfan and Xiao, Qian and Zhao, Tong and Sun, Ninghui and Tan, Guangming and Jia, Weile},
  journal={IEEE Transactions on Computers}, 
  title={29-Billion Atoms Molecular Dynamics Simulation with Ab Initio Accuracy on 35 Million Cores of New Sunway Supercomputer}, 
  year={2025},
  volume={},
  number={},
  pages={1-14},
  url={https://doi.org/10.1109/TC.2025.3540646},
  doi={10.1109/TC.2025.3540646}}

@inproceedings{10.1145/3503221.3508425,
author = {Guo, Zhuoqiang and Lu, Denghui and Yan, Yujin and Hu, Siyu and Liu, Rongrong and Tan, Guangming and Sun, Ninghui and Jiang, Wanrun and Liu, Lijun and Chen, Yixiao and Zhang, Linfeng and Chen, Mohan and Wang, Han and Jia, Weile},
title = {Extending the limit of molecular dynamics with ab initio accuracy to 10 billion atoms},
year = {2022},
isbn = {9781450392044},
publisher = {Association for Computing Machinery},
address = {New York, NY, USA},
url = {https://doi.org/10.1145/3503221.3508425},
doi = {10.1145/3503221.3508425},
booktitle = {Proceedings of the 27th ACM SIGPLAN Symposium on Principles and Practice of Parallel Programming},
pages = {205--218},
numpages = {14},
location = {Seoul, Republic of Korea},
series = {PPoPP '22}
}

@inproceedings{10.1145/3581784.3627041,
author = {Kozinsky, Boris and Musaelian, Albert and Johansson, Anders and Batzner, Simon},
title = {Scaling the Leading Accuracy of Deep Equivariant Models to Biomolecular Simulations of Realistic Size},
year = {2023},
isbn = {9798400701092},
publisher = {Association for Computing Machinery},
address = {New York, NY, USA},
url = {https://doi.org/10.1145/3581784.3627041},
doi = {10.1145/3581784.3627041},
booktitle = {Proceedings of the International Conference for High Performance Computing, Networking, Storage and Analysis},
articleno = {2},
numpages = {12},
location = {Denver, CO, USA},
series = {SC '23}
}

@inproceedings{10.1145/3458817.3487400,
author = {Nguyen-Cong, Kien and Willman, Jonathan T. and Moore, Stan G. and Belonoshko, Anatoly B. and Gayatri, Rahulkumar and Weinberg, Evan and Wood, Mitchell A. and Thompson, Aidan P. and Oleynik, Ivan I.},
title = {Billion atom molecular dynamics simulations of carbon at extreme conditions and experimental time and length scales},
year = {2021},
isbn = {9781450384421},
publisher = {Association for Computing Machinery},
address = {New York, NY, USA},
url = {https://doi.org/10.1145/3458817.3487400},
doi = {10.1145/3458817.3487400},
booktitle = {Proceedings of the International Conference for High Performance Computing, Networking, Storage and Analysis},
articleno = {4},
numpages = {12},
location = {St. Louis, Missouri},
series = {SC '21}
}

@article{Liao2021AscendAS,
  title={Ascend: a Scalable and Unified Architecture for Ubiquitous Deep Neural Network Computing : Industry Track Paper},
  author={Heng Liao and Jiajin Tu and Jing Xia and Hu Liu and Xiping Zhou and Honghui Yuan and Yuxing Hu},
  journal={2021 IEEE International Symposium on High-Performance Computer Architecture (HPCA)},
  year={2021},
  doi={10.1109/hpca51647.2021.00071},
  pages={789-801},
  url={https://api.semanticscholar.org/CorpusID:233376329}
}

@article{dongarra2024hardware,
  title={Hardware trends impacting floating-point computations in scientific applications},
  author={Dongarra, Jack and Gunnels, John and Bayraktar, Harun and Haidar, Azzam and Ernst, Dan},
  journal={arXiv preprint arXiv:2411.12090},
  url={https://arxiv.org/abs/2411.12090},
  year={2024}
}

@article{Bezgin2023,
   author = {Deniz A. Bezgin and Aaron B. Buhendwa and Nikolaus A. Adams},
   doi = {10.1016/j.cpc.2022.108527},
   issn = {00104655},
   journal = {Computer Physics Communications},
   month = {1},
   pages = {108527},
   title = {JAX-Fluids: A fully-differentiable high-order computational fluid dynamics solver for compressible two-phase flows},
   volume = {282},
   url = {https://linkinghub.elsevier.com/retrieve/pii/S0010465522002466},
   year = {2023}
}

@article{Bezgin2025,
   author = {Deniz   Bezgin and Aaron B. Buhendwa and Nikolaus A. Adams},
   doi = {10.1016/j.cpc.2024.109433},
   issn = {00104655},
   journal = {Computer Physics Communications},
   month = {3},
   pages = {109433},
   title = {JAX-Fluids 2.0: Towards HPC for differentiable CFD of compressible two-phase flows},
   volume = {308},
   url = {https://linkinghub.elsevier.com/retrieve/pii/S0010465524003564},
   year = {2025}
}

@article{dongarra2003linpack,
  title={The LINPACK Benchmark: past, present and future},
  author={Dongarra, Jack J and Luszczek, Piotr and Petitet, Antoine},
  journal={Concurrency and Computation: practice and experience},
  volume={15},
  number={9},
  pages={803--820},
  year={2003},
  doi={10.1002/cpe.728},
  url={https://doi.org/10.1002/cpe.728},
  publisher={Wiley Online Library}
}

@inproceedings{haidar2018harnessing,
  title={Harnessing GPU tensor cores for fast FP16 arithmetic to speed up mixed-precision iterative refinement solvers},
  author={Haidar, Azzam and Tomov, Stanimire and Dongarra, Jack and Higham, Nicholas J},
  booktitle={SC18: International Conference for High Performance Computing, Networking, Storage and Analysis},
  pages={603--613},
  year={2018},
  doi={10.1109/sc.2018.00050},
  url={https://doi.org/10.1109/sc.2018.00050},
  organization={IEEE}
}

@inproceedings{lu2022climbing,
  title={Climbing the summit and pushing the frontier of mixed precision benchmarks at extreme scale},
  author={Lu, Hao and Matheson, Michael and Oles, Vladyslav and Ellis, Austin and Joubert, Wayne and Wang, Feiyi},
  booktitle={SC22: International Conference for High Performance Computing, Networking, Storage and Analysis},
  pages={1--15},
  year={2022},
  doi={10.1109/sc41404.2022.00083},
  url={https://doi.org/10.1109/sc41404.2022.00083},
  organization={IEEE}
}

@inproceedings{kudo2020implementation,
  title={Implementation and numerical techniques for one EFlop/s HPL-AI benchmark on Fugaku},
  author={Kudo, Shuhei and Nitadori, Keigo and Ina, Takuya and Imamura, Toshiyuki},
  booktitle={2020 IEEE/ACM 11th Workshop on Latest Advances in Scalable Algorithms for Large-Scale Systems (ScalA)},
  pages={69--76},
  year={2020},
  doi={10.1109/scala51936.2020.00014},
  url={https://doi.org/10.1109/scala51936.2020.00014},
  organization={IEEE}
}

@article{ozaki2012error,
  title={Error-free transformations of matrix multiplication by using fast routines of matrix multiplication and its applications},
  author={Ozaki, Katsuhisa and Ogita, Takeshi and Oishi, Shin’ichi and Rump, Siegfried M},
  journal={Numerical Algorithms},
  volume={59},
  number={1},
  pages={95--118},
  year={2012},
  doi={10.1007/s11075-011-9478-1},
  url={https://doi.org/10.1007/s11075-011-9478-1},
  publisher={Springer}
}

@inproceedings{feng2021egemm,
  title={EGEMM-TC: accelerating scientific computing on tensor cores with extended precision},
  author={Feng, Boyuan and Wang, Yuke and Chen, Guoyang and Zhang, Weifeng and Xie, Yuan and Ding, Yufei},
  booktitle={Proceedings of the 26th ACM SIGPLAN symposium on principles and practice of parallel programming},
  pages={278--291},
  doi={10.1145/3437801.3441599},
  url={https://doi.org/10.1145/3437801.3441599},
  year={2021}
}

@inproceedings{ma2022efficiently,
  title={Efficiently emulating high-bitwidth computation with low-bitwidth hardware},
  author={Ma, Zixuan and Wang, Haojie and Feng, Guanyu and Zhang, Chen and Xie, Lei and He, Jiaao and Chen, Shengqi and Zhai, Jidong},
  booktitle={Proceedings of the 36th ACM International Conference on Supercomputing},
  pages={1--12},
  doi={10.1145/3524059.3532377},
  url={https://doi.org/10.1145/3524059.3532377},
  year={2022}
}

@article{xue2025sgemm,
  title={SGEMM-cube: Emulating FP32 GEMM on Ascend NPUs using FP16 cube units with precision recovery},
  author={Xue, Weicheng and Xu, Baisong and Yang, Kai and Liu, Yongxiang and Fan, Dengdeng and Xu, Pengxiang and Tian, Yonghong},
  journal={arXiv preprint arXiv:2507.23387},
  url={https://arxiv.org/abs/2507.23387},
  year={2025}
}

@book{liang2020ascend,
  title={Ascend AI Processor Architecture and Programming: Principles and Applications of CANN},
  author={Liang, Xiaoyao},
  year={2020},
  doi={10.1016/c2020-0-00270-7},
  url={https://doi.org/10.1016/c2020-0-00270-7},
  publisher={Elsevier}
}

@inproceedings{10.1145/3295500.3356149,
author = {Yang, Kun and Chen, Yi-Fan and Roumpos, Georgios and Colby, Chris and Anderson, John},
title = {High Performance Monte Carlo Simulation of Ising Model on TPU Clusters},
year = {2019},
isbn = {9781450362290},
publisher = {Association for Computing Machinery},
address = {New York, NY, USA},
url = {https://doi.org/10.1145/3295500.3356149},
doi = {10.1145/3295500.3356149},
booktitle = {Proceedings of the International Conference for High Performance Computing, Networking, Storage and Analysis},
articleno = {83},
numpages = {15},
location = {Denver, Colorado},
series = {SC '19}
}

@misc{bayraktar2023cuquantumsdkhighperformancelibrary,
      title={cuQuantum SDK: A High-Performance Library for Accelerating Quantum Science}, 
      author={Harun Bayraktar and Ali Charara and David Clark and Saul Cohen and Timothy Costa and Yao-Lung L. Fang and Yang Gao and Jack Guan and John Gunnels and Azzam Haidar and Andreas Hehn and Markus Hohnerbach and Matthew Jones and Tom Lubowe and Dmitry Lyakh and Shinya Morino and Paul Springer and Sam Stanwyck and Igor Terentyev and Satya Varadhan and Jonathan Wong and Takuma Yamaguchi},
      year={2023},
      eprint={2308.01999},
      archivePrefix={arXiv},
      primaryClass={quant-ph},
      url={https://arxiv.org/abs/2308.01999}, 
}

\section*{Appendix}
\label{artifact_descp}

\subsection{Blas Kernels}
\label{blas_operators}

This section summarizes the BLAS micro-benchmarks used in our study and their mapping to the execution units of Ascend NPUs (Table~\ref{tab:blas_npu_mapping}). We classify each operator according to its computational pattern, numerical precision, and dominant hardware resource, in order to clarify which kernels primarily exercise the Cube engine, the Vector/Scalar pipelines, or both. In addition, we characterize each operator as compute-bound, bandwidth-limited, or latency-dominated based on its arithmetic intensity and memory access behavior. This classification provides a structured interpretation of the performance results reported in Sec.~\ref{sec:core_kernels} and helps explain why different scientific kernels respond differently to the same underlying NPU architecture.

\begin{table}[H]
\centering
\caption{Mapping BLAS Operators to Ascend NPU Execution Units}
\label{tab:blas_npu_mapping}
\resizebox{\textwidth}{!}{
\begin{tabular}{l l c l}
\hline
\textbf{Category} & \textbf{Kernel} & \textbf{Precision} & \textbf{Workload Characteristic} \\
\hline
Dense FP16 GEMM 
& hgemm (+ batched/stride) 
& FP16 
& Cube-dominant, compute-bound \\

Dense FP32 GEMM 
& sgemm (+ batched/stride) 
& FP32 
& Cube + Vector, compute-bound \\

Complex GEMM 
& cgemm (+ batched/stride) 
& Complex FP32 
& Mixed (compute-intensive, Vector-limited) \\

Symmetric Matrix--Matrix 
& ssymm 
& FP32 
& Cube-dominant, structured reuse \\

Symmetric Rank-$k$ Updates 
& ssyrk, ssyr2k 
& FP32 
& Cube-dominant, high reuse \\

Matrix--Vector 
& hgemv, ssymv, sgemv/cgemv (+ batched/stride) 
& FP16 / FP32 / Complex 
& Vector-dominant, memory-bound \\

Triangular Matrix--Vector 
& strmv, ctrmv 
& FP32 / Complex 
& Vector-dominant, latency-bound \\
\hline
\end{tabular}
}
\end{table}

\subsection{SGEMM-cube Application Environment}
The SGEMM-cube application study was conducted on a dedicated single-node testbed.

\begin{itemize}
    \item \textbf{Host Processor (CPU):} The host platform is powered by a high-density Kunpeng 920 ARM processor with a many-core architecture tailored for high throughput.
    \begin{itemize}
        \item \textbf{Core Configuration:} 192 physical cores (64-bit) distributed across 4 physical sockets (48 cores per socket).
        \item \textbf{Topology:} The system is organized into 8 NUMA nodes to optimize local memory access latency, with processes pinned to local NUMA domains during benchmarking to minimize cross-socket traffic.
    \end{itemize}
    \item \textbf{Accelerators (Device):} Comparative evaluations were performed on single-card configurations of two NPU generations:
    \begin{itemize}
        \item \textbf{Target Platform:} Single Ascend 910A NPU.
        \item \textbf{Baseline Platform:} Single Ascend 910B NPU, serving as the next-generation reference.
    \end{itemize}
\end{itemize}

\subsection{LRSVD Experimental Environment}
The Low-Rank SVD (LRSVD) solver was benchmarked on a diverse set of hardware platforms to evaluate performance scaling across NPU generations and to provide a baseline comparison against standard CPU and GPU solvers.

\begin{itemize}
    \item \textbf{Accelerator Platforms (NPU):} Experiments were conducted on single-chip configurations of three successive Ascend generations to demonstrate generational scaling:
    \begin{itemize}
        \item Ascend 910A
        \item Ascend 910B
        \item Ascend 910C
    \end{itemize}
    \item \textbf{Baseline Platform (CPU):} The CPU baseline performance was measured on an Intel Xeon Platinum 8168 processor.
    \item \textbf{Baseline Platform (GPU):}
    The GPU baseline performance was measured on an NVIDIA H800 using cuSOLVER. The plotted configuration uses $r/N=0.1$, two QR refinement loops, and a QR tolerance of 0.005.
\end{itemize}

\subsection{HPL-MxP Testbed Configuration}

\subsubsection{Hardware Environment}
The experimental evaluation is conducted across three distinct hardware platforms to characterize performance at both node-level and cluster-level scales.

\textbf{System A.}
System A is a platform for evaluating large-scale scientific workloads on the Ascend 910A generation. The detailed specification for each node in Cluster A is as follows:
\begin{itemize}
    \item \textbf{Processors (CPU):} Kunpeng 920 CPU (Arm architecture), 192 cores per node.
    \item \textbf{Accelerators (NPU):} 8 $\times$ Ascend 910A NPUs per node.
    \item \textbf{Memory \& Storage:} 2 TB DDR4 memory and 19.2 TB NVMe SSD per node.
    \item \textbf{Interconnect:}
    \begin{itemize}
        \item \textbf{Intra-node:} PCIe 4.0 x16 (8 lanes total, 1 dedicated per NPU).
        \item \textbf{Inter-node:} 100 Gb RoCE v2 network fabric with dedicated interfaces for each NPU (8 $\times$ 100 Gb NICs per node) to ensure non-blocking collective communication.
    \end{itemize}
\end{itemize}

\textbf{System B \& System C.}
To analyze architectural evolution and performance scaling across hardware generations, we utilize two additional platforms:
\begin{itemize}
    \item \textbf{System B (Standalone):} A single-node workstation equipped with 910B accelerators, used exclusively for micro-benchmarking and intra-node architectural characterization.
    \item \textbf{System C (Distributed):} A multi-node cluster representing the latest 910C generation, used to validate weak and strong scaling behavior at scale.
\end{itemize}

\subsubsection{Software Environment}
The software stack on System A is configured to support the full scientific computing pipeline on Ascend 910A architecture:
\begin{itemize}
    \item \textbf{NPU Runtime:} CANN 6.3.RC2
    \item \textbf{Compilers:} Kunpeng GCC 9.3.1, CMake v3.21.4
    \item \textbf{Libraries:} OpenBLAS v0.3.18 (Math), UCX v1.11.2 (Communication)
    \item \textbf{MPI Distribution:} OpenMPI v5.0.0rc2
\end{itemize}

\subsection{PQSim RCS Benchmark}
The inset in Fig.~\ref{fig:PQSim_perf} shows a sketch of the 1D random circuit with qubits number $ N = 7$ and depth $ D = 5$. The gray dotted line outlines the structure of one layer. The circuit is interleaved by layers of single-qubit gates (colored boxes) and two-qubit gates (the dots connected to a blank box). The single-qubit gates are randomly sampled from the set of universal single-qubit quantum gates. The two-qubit gates are either control-NOT or control-Z with equal probability. In performance evaluation of PQSim, $ N = 30 $ and $ D = 30 $ are used to ensure consumed memory matching different chips' configuration and enough entanglement among qubits.

\begin{itemize}
    \item \textbf{Ascend Platforms (NPU):} Experiments were conducted on three successive Ascend generations to demonstrate generational scaling:
    \begin{itemize}
        \item Ascend 910A
        \item Ascend 910B
        \item Ascend 910C: one die is used  
    \end{itemize}
    \item \textbf{Baseline Platform (GPU):}
    The GPU baseline performance was measured on an NVIDIA A800 with cuQuantum 26.01.0.
\end{itemize}

\subsection{SMC-X Experimental Environment}
The SMC-X evaluation uses Ascend 910C as the primary NPU platform. NVIDIA H800 runs the reference SMC-GPU code and provides the high-end scalability baseline for large-scale atomistic simulation, while NVIDIA A800 is used for the kernel-level \texttt{cal\_energy} micro-benchmark.
\end{document}